\documentclass[final,5p,times,twocolumn]{elsarticle}

\usepackage{amsmath}
\usepackage{amssymb}
\usepackage{algorithm}
\usepackage{algorithmic}
\usepackage{booktabs}
\usepackage{multirow}
\usepackage{graphicx}
\usepackage{xcolor}
\usepackage{fancyhdr}
\usepackage{CJKutf8}
\usepackage[hidelinks]{hyperref}

\graphicspath{{figures/}}

\newcommand{\dats}{\textsc{Dats}}
\newcommand{\pass}{\mathrm{pass}}

\definecolor{bannerbg}{HTML}{FFF3CD}
\definecolor{bannerfg}{HTML}{7A4E00}
\definecolor{bannerln}{HTML}{D9B45B}

\journal{Neurocomputing}

\usepackage{etoolbox}
\definecolor{revisionblue}{RGB}{0,0,0}
\makeatletter
\renewcommand\HyOrg@float@makebox[1]{%
  \color@vbox\hsize=#1 \@parboxrestore\color{black}%
    \@fs@pre\@fs@iftopcapt
      \ifvoid\@floatcapt\else\unvbox\@floatcapt\par\@fs@mid\fi
      \unvbox\@currbox
    \else\unvbox\@currbox
      \ifvoid\@floatcapt\else\par\@fs@mid\unvbox\@floatcapt\fi
    \fi\par\@fs@post\vskip\z@\color@endbox}
\makeatother
\newenvironment{blueaddition}{\par\begingroup\color{revisionblue}}{\par\endgroup}
\begin{document}

\begin{frontmatter}

\title{Learning How Much to Collaborate: Difficulty-Aware Topology
Selection for Multi-Agent Code Generation}

\author{Yunsong Hong\corref{cor1}}
\ead{yhon7822@uni.sydney.edu.au}
\cortext[cor1]{Corresponding author.}
\affiliation{organization={School of Computer Science, The University of Sydney},
  addressline={Sydney, NSW 2006}, country={Australia}}

\begin{highlights}
\item Collaboration benefit in multi-agent code generation scales with problem difficulty
\item Topology is treated as a per-problem decision rather than a system-level constant
\item A budget-matched protocol swaps the rank of two baselines once spend is equalised
\item Matched at 40\% of the hierarchical budget, routing adds 4.1 points of pass@1
\item {\color{revisionblue} Math results concern retraining, not unchanged-model transfer}
\end{highlights}

\begin{abstract}

Multi-agent systems for code generation are deployed with a single
{\color{revisionblue} communication topology, chosen once for every problem. Task-dependent allocation is an alternative when collaboration benefits vary across inputs.} Evaluating five topologies on 614 problems from APPS,
HumanEval+ and LiveCodeBench, we find that the advantage of hierarchical
collaboration over a single agent grows from 2.4 points of pass@1 on the easiest
third of problems to 21.1 points on the hardest third, while its token cost stays
about ten times higher. We propose the Difficulty-Aware Topology
Selector (\dats{}), which predicts each topology's probability of solving a
problem and selects the one maximising predicted success minus cost.
Its predictor is a graph network that treats the five topologies as nodes of a
{\color{revisionblue} connectivity order. The complete graph-head comparison has a reported 1.7-point advantage over the flat multi-label head and changes several mechanisms together. Additional capacity-matched adjacency controls show smaller descriptive differences in a separate configuration.} Because the cost penalty is a single scalar recalibrable
{\color{revisionblue} without retraining, routers can be compared at approximately matched spend when the target is attained. With a shared penalty six cost-aware methods span 21.6 percentage points of spend; two leading baselines fall behind DATS after method-specific calibration near the common target. Unmatched rows are reported separately.}
Fixed at 40\% of the always-hierarchical cost, \dats{} reaches 77.7\% pass@1
against 73.6\% (always-hierarchical) and 74.3\% (strongest learned competitor).
{\color{revisionblue} The main comparison includes Holm--Bonferroni-adjusted McNemar tests. Statistical differences at unequal costs do not imply matched-budget superiority.} {\color{revisionblue} Backbone-specific comparisons show gains of 4.07--4.23 points across four backbones. Replacing the 39 interpretable features with a graph}
{\color{revisionblue} network or a pretrained encoder gives a largest observed improvement of 1.3 points among tested alternatives, with no significant pairwise improvement; this does not establish equivalence or an upper bound.} A cross-domain study on 400 mathematical reasoning problems
reproduces the effect, the gap widening from 2.5 to 20.9 points.
\end{abstract}

\begin{keyword}
multi-agent systems \sep large language models \sep code generation \sep
adaptive computation \sep cost-aware routing \sep predict-then-optimize \sep
budget-matched evaluation
\end{keyword}

\end{frontmatter}

\section{Introduction}\label{rev:intro}

\begin{blueaddition}
\paragraph{Practical motivation}
A code-generation service receives tasks with different decomposition and verification needs. Choosing collaboration before execution makes its cost part of resource allocation: the service represents the request, estimates candidate success probabilities, and selects a topology under its cost preference. The candidate communication graph is an execution design; the topology graph inside the predictor is a representation used to compare those designs.

\paragraph{Challenges behind the contributions}
Adaptive allocation requires distinguishing tasks for which collaboration can help from tasks for which it only adds cost. Multi-label prediction accommodates overlapping successful topologies, whereas a single target class discards this overlap. The graph head offers a structured sharing hypothesis, but isolating its benefit requires controls for descriptors, sharing and parameter capacity. Finally, fair cost comparisons require calibration of each method's operating point rather than equality of the penalty coefficient.

\end{blueaddition}

Large language models write code well enough that the interesting engineering
question is no longer whether to use them but how to arrange them. A single
model called once is the cheapest arrangement. Beyond it lies a design space of
multi-agent systems in which several model instances take on roles, exchange
drafts, criticise each other, and converge on an answer. Frameworks such as
ChatDev~\cite{qian2024chatdev}, MetaGPT~\cite{hong2024metagpt} and
AutoGen~\cite{wu2023autogen} have made these arrangements easy to build, and
specialised pipelines such as AgentCoder~\cite{huang2024agentcoder} and
MapCoder~\cite{islam2024mapcoder} report that collaboration lifts pass@1 on
competitive programming benchmarks well above what one call achieves.

What these systems share is a commitment: the topology is fixed at design time.
A pipeline that decomposes, delegates and reviews does so for every incoming
problem, whether the problem asks for a hierarchical decomposition or for a
three-line string manipulation. The commitment is convenient to implement and
expensive to run. In our measurements a hierarchical role-based mesh consumes
9.95 times the tokens of a single call on the same problem; at Claude Opus 4
prices that is \$0.94 versus \$0.09 per problem, a difference that dominates the
operating cost of any deployed system.

The commitment would be defensible if the extra collaboration paid off
uniformly. It does not. Splitting our 614 problems into difficulty tertiles, the
gap between the hierarchical mesh and a single agent is 2.4 points of pass@1 on
the easy third, 10.7 points on the middle third, and 21.1 points on the hard
third. On easy problems the elaborate pipeline spends ten times the budget to
recover two points that a single call very nearly reaches on its own. {\color{revisionblue} An aggregate accuracy gain therefore obscures the different marginal benefits observed across these difficulty strata. Under the cost preferences considered here, additional collaboration offers a smaller accuracy return on the easy stratum.}

{\color{revisionblue} For the heterogeneous task collection and cost preferences studied here, variation in collaboration benefit motivates choosing topology as a function of the input. A fixed topology can remain appropriate for a homogeneous workload or a different cost preference. We therefore ask whether an input-dependent choice can be learned from the problem statement before code generation.}

We show that it can, and that the way the decision is factored matters. The
natural first attempt is to train a classifier that maps a problem directly to
the topology that solved it. This throws away the structure of the decision. The
label ``the topology that solved it'' is ambiguous when several topologies
succeed and undefined when none do, and it hard-codes an accuracy--cost
trade-off into the training targets, {\color{revisionblue} so a classifier trained for a single operating point may need retraining when that point changes. A budget-conditioned classifier is an alternative and is not excluded by this argument.}
Instead we predict what is actually observable---for each topology, whether it
solved each cached problem---and defer the trade-off to a decision rule applied
at inference. The predictor answers five independent questions of the form
``would this topology solve this problem''; the selector combines those answers
with a cost table through a single scalar coefficient. The factorisation is a
{\color{revisionblue} standard predict-then-optimize construction~\cite{elmachtoub2022spo}. The reported difference from direct classification is approximately 6.0 points of pass@1, but the realised budgets differ; this comparison does not isolate an equal-budget causal benefit of factorisation.}

Keeping the trade-off in one scalar has a second consequence that turned out to
matter more than the first, and that we think outlives the router it came from.
{\color{revisionblue} Varying the cost penalty defines a family of routing policies with different attainable operating points; it does not guarantee every spend between the cheapest and most expensive topology,} and reporting its accuracy without pinning its spend says very
little. Giving every method the same coefficient does not repair this, because
the coefficient multiplies each method's own score scale: in our comparison a
shared $\lambda=0.2$ leaves six methods spending anywhere from 30\% to 52\% of
the reference budget. We therefore calibrate every cost-aware method, ours and
the baselines alike, to one common budget before comparing them, choosing each
{\color{revisionblue} method's coefficient by bisection on held-out predictions toward 40\% of the always-hierarchical cost. Attainment is checked separately: four calibrated methods meet the outer tolerance and two do not.} Two baselines that appeared
to beat \dats{} under the shared coefficient fall behind it once their spend is
equalised, and the method that looked best turns out to have been the one
buying the most computation. We state the procedure as a protocol, report what
it does and does not equalise, and regard it as a requirement for this class of
comparison rather than a refinement of it.

The contributions of this work are the following. \emph{First}, we quantify how
the benefit of five collaboration topologies varies with problem difficulty and
{\color{revisionblue} algorithmic type on the 614-problem main evaluation, with additional aggregate comparisons across four backbones. The main evaluation motivates task-dependent allocation.} \emph{Second}, we formulate
topology choice as a per-problem constrained selection problem and instantiate
it as \dats{}, a predict-then-optimize router whose predictor is a compact graph
network over the five candidate topologies: because the topologies form a
connectivity order, we model them as nodes of that order and pass messages along
{\color{revisionblue} it rather than scoring them as independent labels. The combined head comparison has a reported 1.7-point gain at the displayed spend; it does not isolate adjacency. We compare decision heads and input representations separately. Alternatives to the 39 interpretable features have a largest observed improvement of 1.3 points in the tested set, without significant pairwise improvement. This does not prove representation equivalence or locate all gains in the decision head.}
\emph{Third}, we define
the budget-matched evaluation protocol described above, show that it reverses
two of the comparisons among the six cost-aware methods it governs, and use it
throughout. \emph{Fourth}, we
characterise the resulting behaviour and its boundaries: which features carry
the signal, which topologies get chosen for which algorithmic types, how the
router's probabilities calibrate, how much of the gain survives a change of
backbone, and---on 400 GSM8K and MATH problems---whether any of this depends on
the problems being code at all.

\section{Related work}\label{final:related}

\begin{blueaddition}
\paragraph{Routing and cooperation studies}
LLMRouterBench~\cite{li2026llmrouterbench} concerns evaluation of routing among language models; The Avengers~\cite{zhang2026avengers} addresses routing and collective model capabilities. These are useful comparisons for resource allocation and evaluation, while the action here is a communication topology within a fixed catalogue. Work on networking agency~\cite{jia2025agency}, communication under collective risks~\cite{wang2020sentiment}, cognitive context~\cite{wang2018decoy}, identifiable partners~\cite{wang2017onymity}, and regret dynamics~\cite{wang2022regret} provides broader context for interaction design. Evidence from human or social systems does not establish the effectiveness of a language-agent communication graph.

\end{blueaddition}

\subsection{Communication topologies for language agents}

Early multi-agent work with language models established that role assignment and
message passing change what a system can solve.
CAMEL~\cite{li2023camel} paired an instruction-giving agent with an executing
agent; debate procedures~\cite{du2024debate,liang2024debate} ran several
instances against each other and aggregated; generative
agents~\cite{park2023generative} scaled the interaction to societies. For
software specifically, ChatDev~\cite{qian2024chatdev} and
MetaGPT~\cite{hong2024metagpt} encode a waterfall process into agent roles,
AgentVerse~\cite{chen2024agentverse} assembles task-specific teams on demand,
self-collaboration prompting~\cite{dong2024selfcollab} obtains the same effect
from a single model playing several roles, and AutoGen~\cite{wu2023autogen}
provides the conversational substrate on which such processes are assembled.

A second line treats the communication graph itself as an object to optimise.
GPTSwarm~\cite{zhuge2024gptswarm} represents agent systems as graphs and
optimises edges by reinforcement learning; MacNet~\cite{qian2024macnet} studies
how performance scales as the graph grows; AgentPrune~\cite{zhang2025agentprune}
removes redundant edges to cut token spend; G-Designer~\cite{zhang2024gdesigner}
generates a task-conditioned topology with a graph neural network. Our work
shares the premise that topology is a decision variable, and differs in where
the decision is made. These methods search for or generate a graph, usually per
task family, and pay a search or generation cost. We take a small, fixed
catalogue of hand-designed topologies and learn only which catalogue entry to
use for a given problem. The catalogue is coarser, but selection over it costs a
{\color{revisionblue} single forward pass and permits empirical calibration of average spend on validation predictions. This is not a hard budget guarantee on unseen tasks; attainable operating points and realised test costs must be reported.}

Evidence on whether collaboration helps at all is mixed. Wang et
al.~\cite{wang2024bounds} report that multi-agent discussion does not
consistently outperform a single strong agent with a good prompt. Our difficulty
analysis is consistent with both that finding and with the positive reports:
averaged over a benchmark whose problems are mostly easy, discussion looks
unnecessary; restricted to the hard tail, it is decisive.

\subsection{Iterative refinement and self-correction}

Chain, star and mesh topologies inherit their machinery from single-agent
prompting and refinement. Chain-of-thought prompting~\cite{wei2022cot} and
ReAct~\cite{yao2023react} established that structuring a model's intermediate
output changes what it can solve; Self-Refine~\cite{madaan2023selfrefine} and
Reflexion~\cite{shinn2023reflexion} loop generation and critique;
Self-Debug~\cite{chen2024selfdebug} feeds execution feedback back to the model;
Self-Consistency~\cite{wang2023selfconsistency} and Tree of
Thoughts~\cite{yao2023tot} broaden the search rather than deepen the critique.
These are complementary to topology selection: each of our topologies could be
instantiated with a stronger refinement operator, and the routing problem would
remain.

\subsection{Routing and cascades over models}

Choosing which model to call for which query is a well-studied cost-reduction
strategy. FrugalGPT~\cite{chen2023frugalgpt} cascades from cheap to expensive
models with a learned stopping rule; Hybrid LLM~\cite{ding2024hybrid} routes
between a small and a large model on predicted difficulty;
RouteLLM~\cite{ong2024routellm} learns the router from preference data;
ZOOTER~\cite{lu2024zooter} distils reward signals into a routing function;
meta-modelling approaches~\cite{sakota2024flyswat} predict per-model performance
and pick the cheapest adequate one; RouterBench~\cite{hu2024routerbench}
standardises evaluation of such systems. Cascades with mixed thought
representations~\cite{yue2024cascades} apply the same idea to reasoning chains.

Two recent methods route within multi-agent systems rather than over single
models. MasRouter~\cite{zhang2025masrouter} jointly selects agents, models and
collaboration mode; Router-R1~\cite{feng2025routerr1} learns multi-round routing
and aggregation with reinforcement learning. Our formulation is narrower than
either---one decision, made once, over a fixed catalogue---which is what allows
the budget-matched protocol and the significance testing we report. We
reimplement lightweight versions of both as baselines.

The broader question of how much computation to spend per input has been studied
as adaptive computation. Confident Adaptive Language
Modeling~\cite{schuster2022calm} exits early when a per-token confidence
threshold is met; mixture-of-experts routing~\cite{shazeer2017moe} conditions
capacity on the input; test-time compute scaling~\cite{snell2024scaling} and
repeated sampling~\cite{brown2024monkeys} characterise how accuracy responds to
additional inference budget. Topology selection is adaptive computation at the
granularity of an entire agent pipeline.

\subsection{Code generation benchmarks and models}

We evaluate on APPS~\cite{hendrycks2021apps}, which stratifies problems into
introductory, interview and competition levels; HumanEval+, the strengthened
test suite of EvalPlus~\cite{liu2023evalplus} over the original
HumanEval~\cite{chen2021codex}; and LiveCodeBench~\cite{jain2025livecodebench},
which draws from contest problems released after model training cut-offs.
MBPP~\cite{austin2021mbpp}, AlphaCode~\cite{li2022alphacode} and
SWE-bench~\cite{jimenez2024swebench} define adjacent regimes we do not cover.
Our backbones are Claude Opus 4, DeepSeek-V3~\cite{deepseekv3},
GPT-4o-mini~\cite{openai2024gpt4omini} and Qwen2.5-Coder-32B~\cite{hui2024qwen25coder},
the last chosen so that the gradient includes a code-specialised model of the
kind introduced by DeepSeek-Coder~\cite{guo2024deepseekcoder}.

\subsection{Representations of code and of problem statements}\label{rev:related}

A router needs a representation of the input, and the natural candidates come
from code representation learning. CodeBERT~\cite{feng2020codebert} pretrains a
bimodal transformer on code paired with natural language;
GraphCodeBERT~\cite{guo2021graphcodebert} adds data-flow edges to the same
objective; UniXcoder~\cite{guo2022unixcoder} unifies the encoder and decoder
views; code2vec~\cite{alon2019code2vec} predates them with paths through the
abstract syntax tree; and graph convolutional
networks~\cite{kipf2017gcn,velickovic2018gat} supply the machinery for encoding
{\color{revisionblue} either a syntax tree or a statement graph. Program-structure methods require the relevant syntax or data-flow representation. In contrast, pretrained models such as CodeBERT and UniXcoder can encode a natural-language statement without a generated program. Our pre-execution setting excludes unavailable program structure, not natural-language use of pretrained encoders. Section~\ref{sec:encoder} compares selected pretrained encoders and a statement-graph encoder with the hand-crafted features.}

\subsection{Mathematical reasoning as a second domain}

Our cross-domain check uses GSM8K~\cite{cobbe2021gsm8k}, grade-school word
problems with a single numeric answer, and MATH~\cite{hendrycks2021math},
competition problems annotated with five difficulty levels, which is what makes
it usable for a difficulty-scaling analysis. Multi-agent methods have been
applied to both: debate~\cite{du2024debate} and
self-consistency~\cite{wang2023selfconsistency} report gains on arithmetic and
mathematical reasoning, and program-aided
approaches~\cite{gao2023pal} offload the arithmetic to an interpreter. We use
these suites not to compete on them but to ask whether the difficulty-dependent
value of collaboration that we measure on code is a property of code or a
property of hard problems.

\section{Problem formulation}

\begin{figure*}[t]
\centering
\includegraphics[width=\textwidth]{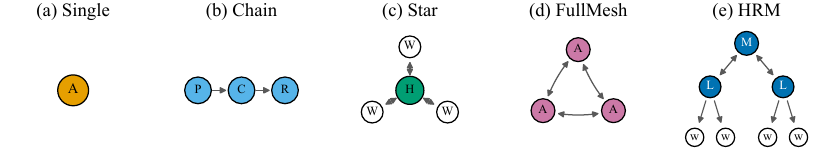}
\caption{The five collaboration topologies. Single issues one call. Chain
passes a draft from proposer (P) to critic (C) to reviser (R). Star has a hub
(H) decompose the problem and delegate to workers (W), whose returns are merged
deterministically.
FullMesh lets three peers (A) read each other's drafts over two rounds. HRM
adds a manager (M) above two leads (L), each supervising two workers (w), with
review flowing upward. Filled nodes indicate agents that see the full problem
statement.}
\label{fig:topologies}
\end{figure*}

Let $x$ denote a problem statement and $\mathcal{T}=\{\tau_1,\dots,\tau_5\}$ the
catalogue of topologies in Fig.~\ref{fig:topologies}. Running $\tau_t$ on $x$
with a fixed backbone yields a program that either passes the problem's full
test suite or does not; write the outcome $y_t(x)\in\{0,1\}$ and its expectation
$s_t(x)=\Pr[\pass\mid x,\tau_t]$. Running $\tau_t$ also consumes tokens, and we
write $c_t$ for the expected cost of topology $t$ in currency units, measured
per topology rather than per problem because the token profile is dominated by
the number and structure of calls rather than by the individual statement.

A router is a map $\pi:\mathcal{X}\to\mathcal{T}$. We want the router that
maximises expected success subject to an average spending constraint,
\begin{equation}
\max_{\pi}\;\mathbb{E}_x\!\left[s_{\pi(x)}(x)\right]
\quad\text{s.t.}\quad
\mathbb{E}_x\!\left[c_{\pi(x)}\right]\le B .
\label{eq:constrained}
\end{equation}
The constraint couples the individual decisions only through an expectation, so
the Lagrangian relaxation separates across problems: for a multiplier
$\lambda\ge0$ the optimal policy is pointwise,
\begin{equation}
\pi_\lambda(x)=\arg\max_{t\in\mathcal{T}}\;\bigl[s_t(x)-\lambda c_t\bigr].
\label{eq:pointwise}
\end{equation}
{\color{revisionblue} For fixed scores and the same fixed topology costs used in selection, $\mathbb{E}_x[c_{\pi_\lambda(x)}]$ is non-increasing in $\lambda$. With a finite catalogue and evaluation set it is generally a step function, so bisection can search for an attainable point near a target but need not attain every target. Only methods meeting the declared tolerance enter a matched-budget comparison. This property concerns the cost used in the objective; it does not by itself guarantee monotonicity of a separately measured per-problem token bill.}

Two quantities in Eq.~\eqref{eq:pointwise} are unavailable at decision time.
The costs $c_t$ are cheap to obtain, since they can be measured once from cached
runs and reused. The success probabilities $s_t(x)$ are the difficult part, and
estimating them is what the next section is about.

Throughout, we call the policy that always plays a single topology
\emph{Always-$\tau$}, and the policy $\pi^{\star}(x)=\arg\max_t y_t(x)$ that
knows the realised outcomes the \emph{oracle}. The oracle is not achievable, but
it bounds what any router built on this catalogue can reach, and the gap to it
measures how much of the available signal a router leaves behind.

\subsection{The five topologies}

\emph{Single} issues one call with the problem statement and a short
instruction. \emph{Chain} runs three sequential calls: a proposer writes a first
solution, a critic lists concrete defects, and a reviser produces the final
program. \emph{Star} uses four calls: a hub decomposes the statement into three
subtasks and three workers solve them independently, after which the returns are
concatenated deterministically into the final program without a further model
call.
\emph{FullMesh} runs three peers for two rounds, six calls in total, with each
peer seeing the other two drafts in the second round. \emph{HRM}, the
hierarchical role-based mesh, places a manager above two leads, each of which
supervises two workers; leads consolidate their workers' output and the manager
arbitrates, for five to eight calls depending on whether the manager requests a
second pass. The catalogue is deliberately small and hand-designed: our claim
concerns the value of choosing among topologies, not the value of any particular
topology.

\section{Difficulty-aware topology selection}

\begin{figure*}[t]
\centering
\includegraphics[width=\textwidth]{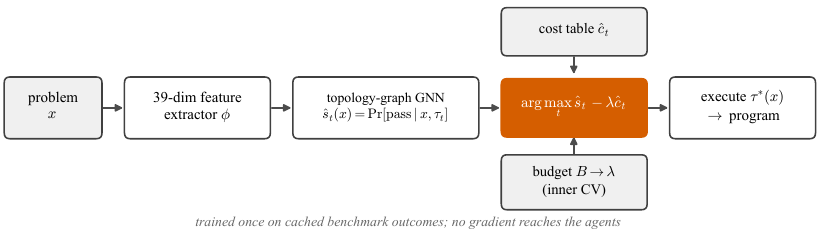}
\caption{\dats{}. A problem statement is mapped to 39 features and a shared
trunk; a graph network then passes messages over the five candidate topologies,
arranged by connectivity, and reads out one success probability per topology.
The selector subtracts a cost term whose coefficient is calibrated to the target
budget on held-out data. Only the selected topology is executed.}
\label{fig:framework}
\end{figure*}

\dats{} estimates $s_t(x)$ for all five topologies from the problem statement
alone, then applies Eq.~\eqref{eq:pointwise} with a calibrated multiplier
(Fig.~\ref{fig:framework}). Nothing in the pipeline requires executing a
topology at decision time, and nothing propagates gradients into the agents.

\begin{blueaddition}
\paragraph{Prediction and selection}
The predictor estimates candidate-specific success scores from the same task representation. The selection rule then compares these scores after the cost penalty. This separation allows the preferred action to change with the deployment budget without redefining a unique correct topology for every training example. A graph-based head can share information between candidate descriptions; such sharing is a modelling hypothesis, not by itself evidence of greater predictive accuracy. Flat, node-wise, descriptor and adjacency controls address different parts of that hypothesis.

\end{blueaddition}

\subsection{Features}\label{final:features}

The feature map $\phi:\mathcal{X}\to\mathbb{R}^{39}$ is deliberately
interpretable, because we want to be able to say afterwards which properties of
a problem drive the decision. It has six groups.

The \emph{lexical} group (five dimensions) records length in words, sentence
count, average word length, the fraction of tokens that appear inside code
spans, and a constraint density obtained by counting phrases such as
``at most'', ``exactly'' and ``for each''. The \emph{structural} group (four)
counts worked examples, detects an explicit edge-case discussion, counts
required functions in the signature, and scores input--output complexity from
the described formats. The \emph{complexity} group (three) captures cues that a
solution will need nesting or asymptotic care: depth hints such as ``for each
pair'', the number of large numeric bounds, and the count of algorithmic
keywords associated with hard techniques. The \emph{difficulty} group is a
{\color{revisionblue} single scalar predicted from the statement. A ridge regressor with an $L_2$ penalty of 1 uses the twelve lexical, structural and complexity features to predict a five-level difficulty rating. Training ratings follow a fixed rubric: direct translation, one elementary technique, composition of familiar techniques, an advanced technique, and several interacting advanced techniques. The prediction is clipped to $[1,5]$ and mapped to $[0,1]$ by $(r-1)/4$. The scorer and its input standardisation are fitted within each training fold; training examples receive inner out-of-fold difficulty predictions. Validation and test ratings are never supplied to the router. No language-model call is needed. The same scoring procedure is used in the primary and additional comparisons.} The \emph{semantic} group (sixteen) holds the leading singular
directions of a TF--IDF representation of the statement. The
\emph{algorithmic} group (ten) is a one-hot encoding of the problem's dominant
technique, assigned by keyword vote over ten categories from implementation and
string handling to dynamic programming and graphs; the vote recovers our manual
labels on 85.5\% of problems.

The two groups that a router might be expected to lean on hardest, difficulty
and algorithmic type, are also the two that carry the most signal in the
ablation of Section~\ref{sec:ablation}. Their interaction is what makes the
problem non-trivial: the useful pattern is not ``hard problems need HRM'' but
``hard \emph{graph} problems need HRM while hard \emph{implementation} problems
do not'', and a model without interaction terms cannot express it.

\subsection{Predictor}\label{final:predictor}

{\color{revisionblue} The predictor maps a task representation to five marginal success probabilities. A flat multi-label predictor is a valid baseline; separate sigmoid losses do not by themselves establish statistical independence of outcomes. The proposed topology graph expresses a hypothesis that structural descriptors and shared computation can help predict correlated candidate outcomes. The connectivity order does not require success to be nested: a heavier topology can fail on a problem solved by a lighter one. Whether a particular adjacency helps must be tested against controlled alternatives.}

We therefore make the label space explicit and predict over it with a small
graph network. Let $G_{\mathcal{T}}=(\mathcal{T},E)$ be the \emph{topology
graph}: its five nodes are the topologies and its edges join each topology to its
neighbours in the connectivity order, with the hierarchical mesh linked to Star
and FullMesh as the two structures it composes. Each node carries a static
descriptor $e_t\in\mathbb{R}^{4}$---agent count, edge count, diameter and mean
degree of the \emph{topology's own} communication graph---so the network is told
what each option is, not merely that five options exist.

{\color{revisionblue} The problem enters through a shared trunk with affine layers of widths 128 and 64, each followed by a ReLU. The 39 standardised features are mapped to a task context}
{\color{revisionblue} $z=\mathrm{trunk}(\phi(x))\in\mathbb{R}^{64}$, and each node is initialised by}
concatenating that context with its structural descriptor,
\begin{equation}
h_t^{0}=\mathrm{ReLU}\!\bigl(W_{0}\,[\,z\;\Vert\;e_t\,]+b_{0}\bigr).
\label{eq:nodeinit}
\end{equation}
The trunk is shared across nodes for the same reason a trunk was shared across
the old heads---difficulty is a property of the problem, not of the topology---but
now the topology-specific part of the computation is a message passing over
$G_{\mathcal{T}}$ rather than five detached linear maps. For $\ell=0,\dots,L-1$,
\begin{equation}\color{revisionblue}
h_t^{\ell+1}=\mathrm{ReLU}\!\Bigl(W_{\mathrm{self}}^{\ell}\,h_t^{\ell}
+W_{\mathrm{nbr}}^{\ell}\!\!\sum_{u\in\mathcal{N}(t)}\!\!\tfrac{1}{|\mathcal{N}(t)|}\,h_u^{\ell}+b^{\ell}\Bigr),
\label{eq:mp}
\end{equation}
{\color{revisionblue} with weights and biases shared across nodes, separate parameters in each of the $L=2$ rounds, and four-dimensional node states in the primary configuration. Thus a topology's score is a}
function of the predicted evidence at the topologies structurally adjacent to it.
A per-node sigmoid readout $\hat{s}_t=\sigma(w^{\top}h_t^{L}+b)$ produces the five
probabilities. {\color{revisionblue} The primary neural predictor has $5,120+8,256+276+72+5=13,729$ parameters, reported as 13.7K: two trunk layers, a 68-to-4 node initialiser, two message-passing rounds and a shared readout. The scalar difficulty regressor is a separate feature-extraction component. Section~\ref{rev:graph-controls} additionally uses a 64-unit trunk and 32-unit node states, with 8,961 neural parameters; all adjacency variants in that comparison have identical capacity.} Inputs are
standardised on the training fold; optimisation uses Adam at an initial learning
rate of $10^{-3}$ with early stopping on a 10\% internal validation split and a
cap of 400 epochs. The loss is the mean binary cross-entropy over the five node
readouts,
\begin{equation}
\mathcal{L}(\theta)=-\frac{1}{5N}\sum_{i=1}^{N}\sum_{t=1}^{5}
\Bigl[y_t^{(i)}\log \hat{s}_t^{(i)}+(1-y_t^{(i)})\log(1-\hat{s}_t^{(i)})\Bigr].
\label{eq:loss}
\end{equation}

\begin{figure*}[t]
\centering
\includegraphics[width=\textwidth]{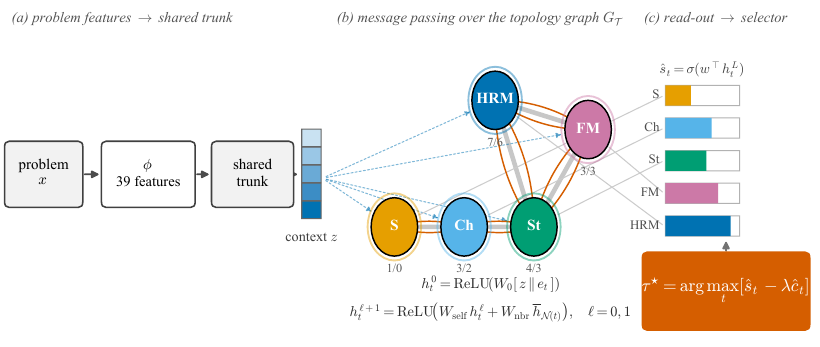}
\caption{The topology-graph predictor. \textbf{(a)}~A problem $x$ is mapped by
$\phi$ to 39 difficulty features and a shared trunk to a context
$z\in\mathbb{R}^{d}$. \textbf{(b)}~The five candidate topologies are the nodes of
the topology graph $G_{\mathcal{T}}$, joined along the connectivity order
Single--Chain--Star--FullMesh with the hierarchical mesh composing Star and
FullMesh; each node is initialised from $z$ (dashed) and its own structural
descriptor $e_t$ (agent/edge counts shown), then exchanges messages with its
structural neighbours (orange) for $L=2$ rounds. \textbf{(c)}~A per-node sigmoid
readout produces one success probability per topology, which the cost-aware
selector ranks. The graph the network reasons over is the \emph{output} space of
topologies, not the absent program graph of the input. Bias terms and round superscripts are omitted from the schematic for clarity; the complete update is given in Eq.~\eqref{eq:mp}.}
\label{fig:gnn}
\end{figure*}

{\color{revisionblue} The input representation and output head are distinct design choices. A program syntax graph is unavailable before code generation; a natural-language statement graph and text encoders are available. The output topology graph is a predefined relation between candidate communication designs. Section~\ref{sec:ablation} compares the complete head with a flat alternative, without isolating adjacency. Figure~\ref{fig:gnn} illustrates the shared trunk, message passing and per-node readout; the additional controlled configuration is specified in Section~\ref{rev:graph-controls}.}

Two design points carry over unchanged because they are easy to get wrong.
First, the targets are per-topology outcomes, not a single categorical label, so
problems that every topology solves and problems that none solve both
{\color{revisionblue} contribute to the binary prediction loss. For example, targets $(1,0,1,0,0)$ retain two successful options, while $(0,0,0,0,0)$ trains all success scores downward without inventing a successful class. An all-zero observed target does not prove every future rollout will fail. The selector favours the cheapest option only when predicted success scores are tied (with a positive cost penalty); that choice is not an extra training label.} Second,
the network never sees cost, which is what keeps the trade-off confined to the
selector.

\subsection{Selection and budget calibration}\label{final:selection}

Costs enter through a normalised table $\hat{c}_t=c_t/\max_{t'}c_{t'}$, so
$\hat{c}_{\text{Single}}=0.10$ and $\hat{c}_{\text{HRM}}=1$. The selector is
\begin{equation}
\tau^{\star}(x)=\arg\max_{t\in\mathcal{T}}
\bigl[\hat{s}_t(x)-\lambda\,\hat{c}_t\bigr],
\label{eq:select}
\end{equation}
which places $\lambda$ on the same scale as a probability difference: at
$\lambda=0.06$, upgrading from Single to HRM must buy at least 5.4 points of
predicted success to be worth it.

The multiplier is not a hyperparameter to be tuned for accuracy. It is
{\color{revisionblue} determined by the budget. The mean monetary target is $B=\beta c_{\rm HRM}$, where $\beta$ is the dimensionless HRM-relative fraction reported in the tables. Normalising costs in Eq.~\eqref{eq:select} rescales the multiplier, not the budget. We run bisection on $\lambda$ using three-fold inner predictions, evaluating the selected fixed-table mean cost at each step, and stop}
{\color{revisionblue} when the inner realised spend is within 0.5 percentage points of the Always-HRM reference cost from the target $B$, or after sixty iterations. An unattainable target is reported as unmatched rather than guaranteed by the search.} The
resulting $\lambda$ is then applied unchanged to the outer test fold.
Algorithm~\ref{alg:dats} states the procedure. Across the five outer folds of
our main experiment the calibrated values were 0.058, 0.052, 0.065, 0.056 and
0.061, a spread small enough that the routing behaviour is stable across folds
but large enough that fixing a single value in advance would miss the budget.

\begin{algorithm}[H]
\small\color{revisionblue}
\caption{\textcolor{revisionblue}{Nested training and budget calibration}}
\label{alg:dats}
\begin{algorithmic}[1]
\REQUIRE outer training set $\mathcal D_{\rm tr}$, untouched outer test set $\mathcal D_{\rm te}$, cached training outcomes, fixed monetary cost table $c$, target $B=\beta c_{\rm HRM}$, HRM reference cost $c_{\rm HRM}$
\STATE Split $\mathcal D_{\rm tr}$ into three inner folds; allocate inner out-of-fold scores $S$
\FOR{each inner validation fold $V$}
  \STATE Set $T=\mathcal D_{\rm tr}\setminus V$
  \STATE Fit preprocessing and the difficulty scorer on $T$; use cross-fitted difficulty scores within $T$
  \STATE Fit predictor on $T$; early stopping uses a split contained in $T$
  \STATE Transform and score $V$ with the frozen inner pipeline; store predictions in $S_V$
\ENDFOR
\STATE Set $\lambda_{\rm lo}=0$, $\lambda_{\rm hi}=2$; evaluate attainable endpoint costs using $S$
\IF{$B$ is outside the attainable endpoint range}
  \STATE Mark the target unmatched and retain the nearest endpoint
\ELSE
  \FOR{$j=1$ \TO $60$}
    \STATE $\lambda\leftarrow(\lambda_{\rm lo}+\lambda_{\rm hi})/2$
    \STATE Select by $S-\lambda c/c_{\rm HRM}$; compute mean selected monetary cost using $c$
    \STATE Retain the candidate with smallest absolute target error
    \IF{absolute target error is at most $0.005\,c_{\rm HRM}$}
      \STATE \textbf{break} the bisection loop
    \ENDIF
    \IF{cost exceeds $B$} \STATE $\lambda_{\rm lo}\leftarrow\lambda$
    \ELSE \STATE $\lambda_{\rm hi}\leftarrow\lambda$ \ENDIF
  \ENDFOR
\ENDIF
\STATE Select the retained multiplier; mark unmatched if error exceeds $0.5\%$ of Always-HRM reference cost
\STATE Refit preprocessing and predictor within $\mathcal D_{\rm tr}$ using the selected training specification; freeze the pipeline and multiplier
\STATE Evaluate $\mathcal D_{\rm te}$ once; report accuracy and realised cost separately from the target
\ENSURE fitted predictor and multiplier; no outer-test tuning
\end{algorithmic}
\end{algorithm}

\begin{blueaddition}
\paragraph{Training and validation isolation}
For each outer split, the difficulty scorer, TF--IDF vocabulary, SVD projection, standardisation and early stopping are fitted exclusively within the training portion. Inner validation predictions come from a pipeline that has not fitted those examples. We choose the multiplier from the pooled inner out-of-fold scores, refit the pipeline on the outer training portion, and evaluate the untouched test fold. Algorithm~\ref{alg:dats} is used for both primary and additional comparisons. Section~\ref{rev:multibudget} specifies the additional inner-search refinement used for the recalibrated baseline points. A refit can change score calibration, so outer cost is checked independently against the declared tolerance.

\end{blueaddition}

\subsection{The budget-matched protocol}
\label{sec:protocol}

The calibration step above was introduced as a component of \dats{}. We now
separate it from \dats{}, because it is not specific to our router and because
without it the accuracies that this class of method reports are not comparable
to each other. We state it as a protocol and treat it as a contribution
independent of the model it was developed for.

The difficulty it addresses is that a cost-aware router is not a system but a
one-parameter family of systems. Any method of the form
``score minus $\lambda$ times cost'' sweeps out a curve in the cost--accuracy
plane as $\lambda$ moves, and every point of that curve is a legitimate
configuration of the same method. An accuracy reported without its realised
spend therefore identifies a curve only up to an unknown position along it, and
two such numbers cannot be subtracted. The obvious repair---give every method
the same $\lambda$---does not work, because $\lambda$ acts on each method's own
score scale. A predictor whose probabilities are compressed towards $0.5$
upgrades to an expensive topology less often, at identical $\lambda$, than one
whose probabilities are spread towards the extremes, so a shared coefficient
silently places the methods at different operating points. In our comparison a
shared $\lambda=0.2$ spreads realised spend over 21.6 percentage points of the
reference budget (Section~\ref{sec:whatmatching}).

The protocol has four steps. \emph{One}, declare a budget $B$ as a fraction of
the realised spend of a reference policy that requires no calibration; we use
the most expensive fixed topology, which makes $B$ interpretable as ``this
fraction of what the standard pipeline costs''. \emph{Two}, calibrate each
method separately, by bisection on its own coefficient inside the training
folds, never on the test fold. \emph{Three}, verify the realised spend out of
fold, report it beside the accuracy, and declare a method budget-matched only
{\color{revisionblue} if it lands inside the stated outer-evaluation tolerance of 0.4 percentage points. Four of the six calibrated methods meet it; E2E and Difficulty-Threshold do not. The inner search uses its separately stated 0.5-percentage-point tolerance.} \emph{Four}, compare accuracies only within the matched
group, and report policies that cannot be calibrated---fixed topologies, the
oracle---separately, as locators of the frontier rather than as competitors.

Three properties make the protocol cheap to adopt. It requires only that
{\color{revisionblue} the cost used for the search be monotone in the coefficient. This holds for fixed scores when the same fixed cost table is used in both selection and search evaluation: raising $\lambda$ cannot increase that selected cost. A separately calculated token bill requires its own empirical check.} It
converges quickly, needing 21.4 bisection steps on average to reach the $0.5\%$
inner tolerance. And on a cached evaluation grid it costs no inference at all,
since every candidate assignment is scored against outcomes that have already
been paid for. What it does not equalise should also be stated: it matches
expected monetary spend, not latency, not worst-case cost, and not the variance
of either. A method that meets the budget by occasionally choosing a very
expensive topology is matched to one that never does, and the two may be quite
different to operate.

\subsection{Cost of the router}\label{rev:cost}

{\color{revisionblue} The 39--128--64--5 shorthand specifies the input, trunk widths and five output scores; the intervening four-dimensional node states are specified in Section~4.2. The primary router has 13.7K neural parameters and takes 28 seconds to train on 614 problems on CPU. Feature extraction, including the linear difficulty scorer, takes under a millisecond per problem. The cheapest language-model topology takes 8.4 seconds. Monetary budgets account for topology execution tokens; local feature extraction and router computation are reported separately and do not require a paid model call.}

\section{Experimental setup}

\subsection{Problems}

We evaluate on 614 problems (Table~\ref{tab:data}): 300 from APPS, one hundred
at each of the introductory, interview and competition levels; the full 164
problems of HumanEval+; and 150 problems from LiveCodeBench drawn from contests
held after the training cut-off of every backbone we test. The suites were
chosen to span difficulty rather than to maximise size: HumanEval+ supplies a
regime where a single call almost always suffices, APPS-Competition supplies one
where nothing reliably works, and the middle suites populate the interval
between them where routing decisions are actually contested. Mean difficulty in
Table~\ref{tab:data} is the normalised rating described in Section~4.1,
reported here to show that the suites are ordered as intended.

\begin{table}[t]
\centering
\small
\setlength{\tabcolsep}{4pt}
\caption{Composition of the evaluation set. Difficulty is the normalised
statement rating; words and tests are per-problem means. {\color{revisionblue} Tests counts evaluator groups, each of which can contain multiple input cases or assertions.}}
\label{tab:data}
\begin{tabular}{lrrrr}
\toprule
Suite & $n$ & Difficulty & Words & Tests \\
\midrule
APPS-Introductory & 100 & 0.25 & 178 & 12.4 \\
APPS-Interview    & 100 & 0.50 & 296 & 18.7 \\
APPS-Competition  & 100 & 0.75 & 412 & 21.3 \\
HumanEval+        & 164 & 0.20 &  67 & 24.6 \\
LiveCodeBench     & 150 & 0.59 & 341 & 16.2 \\
\midrule
Total             & 614 & 0.44 & 232 & 19.0 \\
\bottomrule
\end{tabular}
\end{table}

\subsection{Backbones and execution}

Claude Opus 4 is the primary backbone; DeepSeek-V3, GPT-4o-mini and
Qwen2.5-Coder-32B provide a capability gradient for the generalisation study.
Every agent inside a topology uses the same backbone, so the comparison isolates
topology from model choice. Decoding uses temperature 0.2 and a 4096-token
output cap, and we report pass@1 from a single sample, since a router that only
helps under repeated sampling would be solving a different problem. A program counts as correct only if it
passes the complete test suite. {\color{revisionblue} The evaluator executes all configured checks for the problem. The Tests column counts grouped checks in the execution harness, rather than the number of raw inputs in an expanded benchmark distribution; in particular, it must not be read as the raw HumanEval+ assertion count.}

The full grid is $614\times5\times4=12{,}280$ executions. All outcomes, token
counts and generated programs are cached, so the routing experiments---five
folds times five seeds times thirteen methods---read from the cache and incur no
further inference cost. Total spend for the grid was approximately \$1{,}807, of
which \$1{,}753 was Claude Opus 4.

\subsection{Cost accounting}

Table~\ref{tab:cost} reports the measured token profile. Costs use published
per-token prices; for Claude Opus 4 we combine \$15 per million input tokens and
\$75 per million output tokens, which given our observed input--output mix
yields an effective \$36 per million. The ratios in the fourth column are the
quantity that matters for routing, and they are stable across backbones to
within three percent, because they are determined by the number and structure of
calls rather than by the model. Latency is wall-clock median over the grid, with
independent workers in Star and FullMesh executed in parallel.

\begin{table}[t]
\centering
\small
\setlength{\tabcolsep}{4pt}
\caption{Measured cost profile per topology on Claude Opus 4. Tokens are
per-problem means over the full grid; ratio is relative to Single.}
\label{tab:cost}
\begin{tabular}{lrrrrr}
\toprule
Topology & Calls & Tokens & Ratio & USD & Lat.\,(s) \\
\midrule
Single   & 1     &  2{,}632 & 1.00 & 0.095 &  8.4 \\
Chain    & 3     & 13{,}425 & 5.10 & 0.483 & 26.1 \\
Star     & 4     & 14{,}083 & 5.35 & 0.507 & 21.7 \\
FullMesh & 6     & 23{,}004 & 8.74 & 0.828 & 44.3 \\
HRM      & 5--8  & 26{,}189 & 9.95 & 0.943 & 38.6 \\
\midrule
\dats{}  & 2.6   & 10{,}476 & 3.98 & 0.377 & 17.1 \\
\bottomrule
\end{tabular}
\end{table}

\subsection{Protocol}\label{final:protocol}

All routers are evaluated by five-fold cross-validation stratified jointly on
suite and difficulty tertile, repeated over five random seeds; reported accuracy
is the pooled out-of-fold accuracy of the first seed, and the seed-to-seed
standard deviation is reported alongside. Hyperparameters and the cost
multiplier are selected inside a three-fold split of each training fold and
never touch the corresponding test fold.

The budget is fixed at $B=40\%$ of the always-hierarchical cost, that is
{\color{revisionblue} \$0.377 per problem as the nominal target. Calibrated methods use their own coefficients and searches under Section~\ref{sec:protocol}; only those whose reported outer cost meets the declared tolerance enter matched-budget ranking.} Fixed topologies obviously cannot be calibrated and are reported
at their natural cost; they are included to locate the frontier, not as
budget-matched competitors.

\subsection{Baselines}\label{final:baselines}

Five \emph{Always-$\tau$} policies establish the fixed-topology frontier.
\emph{Random} samples a topology uniformly and is included to confirm that the
routing gain is not an artefact of mixing. \emph{Difficulty-Threshold} bins the
predicted difficulty rating and maps bins to topologies, with thresholds set on
the training fold to meet the budget; it represents the intuition that
difficulty alone should be enough. {\color{revisionblue} DATS-E2E is a direct five-way classification reference. Its stated target is the cheapest successful topology, which is defined only when at least one candidate succeeds. The retained oracle count implies $614-546=68$ problems with no successful topology in the cached grid. All-failed training examples receive the cheapest-topology fallback class, retaining all training examples. Ties between successful candidates use the same cost order. The softmax baseline is trained with cross-entropy; this explicit fallback rule is used throughout the E2E comparisons. Its displayed accuracy is reported at its actual operating cost, not as a controlled isolation of predict-then-optimize factorisation. The fallback class is an explicit single-label convention, distinct from the multi-label all-zero target in Section~4.2.} \emph{MasRouter-lite} follows
\cite{zhang2025masrouter} in learning a routing distribution over collaboration
modes, restricted here to a logistic model on the lexical and structural
features. \emph{RouterR1-lite} follows \cite{feng2025routerr1} in learning a
reward regressor---ridge regression on accuracy minus normalised cost---and
selecting the argmax. \emph{RandomForest-Router} substitutes a 300-tree forest
for the network in the same predict-then-optimize scaffold and is the strongest
learned competitor. The \emph{oracle} completes the table.

\subsection{Statistics}

Accuracy intervals are Wilson score intervals at 95\%~\cite{wilson1927}. Paired comparisons
between routers on the same problems use McNemar's test~\cite{mcnemar1947},
exact binomial when the number of discordant pairs is below 25 and with Yates'
continuity correction otherwise, as recommended for classifier comparison by
Dietterich~\cite{dietterich1998tests}. Families of comparisons are corrected by
the Holm--Bonferroni procedure~\cite{holm1979}. {\color{revisionblue} The six feature-group ablations form one comparison family; their significance markers use Holm-adjusted McNemar $p$ values.} Calibration is summarised by
expected calibration error over ten equal-width bins~\cite{naeini2015ece,guo2017calibration}
and by the Brier score~\cite{brier1950}. {\color{revisionblue} Scikit-learn~\cite{pedregosa2011sklearn} implements ridge scoring, TF--IDF/SVD preprocessing, random forests and the linear probes. PyTorch implements the topology head and statement GCN; transformer encoders use the Transformers interface with PyTorch. The graph models use Adam, batch size 64, an initial learning rate of $10^{-3}$, at most 400 epochs and early stopping with patience 30 on training-side validation loss. Seeds 0--4 control initialisation and minibatch order; data partitions are held fixed for paired model comparisons.}

\section{Results}

\subsection{No topology dominates}

Table~\ref{tab:topo} reports pass@1 for the five topologies on Claude Opus 4.
The ordering by aggregate accuracy is Single, Star, Chain, FullMesh, HRM, and
the total spread is 11.4 points. Two features of the table matter more than the
ordering.

The first is how unevenly the spread is distributed. On HumanEval+ the five
topologies lie within 4.2 points of each other, and the cheapest is only 4.2
points behind the most expensive despite costing a tenth as much. On
APPS-Competition the same comparison is 22 points. Any system that fixes one
topology is implicitly accepting one of these two trade-offs everywhere.

The second is that the ordering is not a ranking of collaboration intensity.
Star uses four calls and Chain three, yet Chain is ahead of Star on every suite.
Sequential criticism, where each agent sees the full statement and the previous
draft, extracts more from three calls than decomposition and integration extract
from four, because decomposition can split a problem along the wrong seam and
neither the workers nor the hub can recover from that. The pairwise tests in
\ref{app:pairwise} make the same point statistically: Chain and FullMesh are
indistinguishable ($p=0.628$) although FullMesh costs 71\% more.

\begin{table}[t]
\centering
\small
\setlength{\tabcolsep}{3.5pt}
\caption{pass@1 (\%) by topology and suite on Claude Opus 4. The last row gives
the number of solved problems out of 614.}
\label{tab:topo}
\begin{tabular}{lrrrrr}
\toprule
Suite & Single & Chain & Star & FullMesh & HRM \\
\midrule
APPS-Introductory & 88.0 & 91.0 & 89.0 & 90.0 & \textbf{92.0} \\
APPS-Interview    & 57.0 & 66.0 & 63.0 & 67.0 & \textbf{71.0} \\
APPS-Competition  & 19.0 & 32.0 & 27.0 & 35.0 & \textbf{41.0} \\
HumanEval+        & 90.9 & 93.3 & 92.1 & 94.5 & \textbf{95.1} \\
LiveCodeBench     & 46.0 & 56.0 & 52.7 & 57.3 & \textbf{61.3} \\
\midrule
Overall           & 62.21 & 69.38 & 66.61 & 70.52 & \textbf{73.62} \\
Solved            & 382 & 426 & 409 & 433 & \textbf{452} \\
\bottomrule
\end{tabular}
\end{table}

\subsection{The benefit of collaboration scales with difficulty}\label{final:difficulty}

\begin{figure}[t]
\centering
\includegraphics[width=\columnwidth]{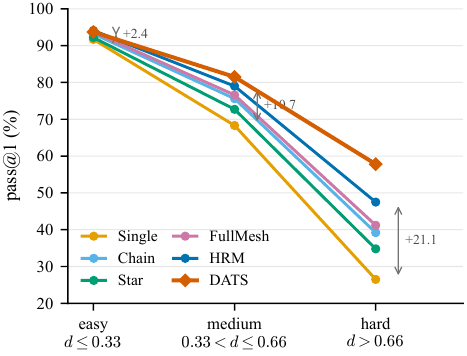}
\caption{pass@1 by difficulty tertile. Arrows mark the HRM-minus-Single gap,
which widens from 2.4 to 21.1 points. \dats{} exceeds every fixed topology on
the hard tertile while spending 40\% of the HRM budget overall.}
\label{fig:difficulty}
\end{figure}

Fig.~\ref{fig:difficulty} splits the 614 problems into difficulty tertiles. The
five topology curves fan out as difficulty rises: on the easy third they occupy
a 2.4-point band, on the middle third 10.7 points, on the hard third 21.1
points. The hierarchical mesh is not uniformly better than a single call; it is
barely better on easy problems and decisively better on hard ones, at constant
relative cost.

{\color{revisionblue} This reported heterogeneity motivates task-dependent allocation. A fixed topology can remain appropriate for a homogeneous workload or a particular cost preference; these aggregate strata do not establish misallocation on every problem.} A
system fixed on Single forgoes twenty-one points on the hard tail. A system
fixed on HRM spends ten times the budget on the easy two-thirds to gain between
two and ten points, and in the easy third specifically it spends \$0.94 per
problem to move 91.7\% to 94.1\%.

The \dats{} curve in the same figure shows what per-problem allocation buys. It
tracks the hierarchical mesh on the easy third, where it mostly declines to pay
for collaboration, and exceeds it by 10.3 points on the hard third, where it
concentrates the budget it saved. The router does not merely interpolate between
the fixed policies; on the hard tertile it is above all five of them, because
different hard problems want different topologies and no single policy can
follow.

\subsection{Budget-matched comparison}

{\color{revisionblue} Table~\ref{tab:main} reports the original accuracy and realised cost of each method. The nominal target for calibrated routers is \$0.377 per problem, or 40\% of Always-HRM. DATS, RandomForest-Router, RouterR1-lite and MasRouter-lite are within 0.4 percentage points of that target. DATS-E2E (38.3\%) and Difficulty-Threshold (39.3\%) fall outside the stated 0.4-percentage-point outer-evaluation tolerance; Random (60.6\%) is also outside it. These rows remain reference operating points and are excluded from matched-budget ranking.}

\begin{table*}[t]
\centering
\small
\setlength{\tabcolsep}{6pt}
\caption{Budget-matched comparison on Claude Opus 4, 614 problems, five-fold
cross-validation. Learned routers are calibrated to $B=40\%$ of the
{\color{revisionblue} always-hierarchical cost as a nominal target. Realised costs determine matching: E2E, Difficulty-Threshold and Random are unmatched references. Fixed topologies are shown at natural cost. Bold identifies the best accuracy among the matched routers.}}
\label{tab:main}
\begin{tabular}{llrrcrr}
\toprule
& Method & Solved & pass@1 (\%) & 95\% CI & USD/problem & \% of HRM cost \\
\midrule
\multirow{5}{*}{\rotatebox{90}{fixed}}
& Always-Single       & 382 & 62.21 & [58.3,\,66.0] & 0.095 & 10.1 \\
& Always-Chain        & 426 & 69.38 & [65.6,\,72.9] & 0.483 & 51.3 \\
& Always-Star         & 409 & 66.61 & [62.8,\,70.2] & 0.507 & 53.8 \\
& Always-FullMesh     & 433 & 70.52 & [66.8,\,74.0] & 0.828 & 87.8 \\
& Always-HRM          & 452 & 73.62 & [70.0,\,77.0] & 0.943 & 100.0 \\
\midrule
\multirow{7}{*}{\rotatebox{90}{\textcolor{revisionblue}{routing policies}}}
& Random              & 421 & 68.57 & [64.8,\,72.1] & 0.571 & 60.6 \\
& Difficulty-Threshold& 432 & 70.36 & [66.6,\,73.8] & 0.371 & 39.3 \\
& \dats{}-E2E         & 440 & 71.66 & [68.0,\,75.1] & 0.361 & 38.3 \\
& MasRouter-lite      & 446 & 72.64 & [69.0,\,76.0] & 0.375 & 39.8 \\
& RouterR1-lite       & 451 & 73.45 & [69.8,\,76.8] & 0.376 & 39.9 \\
& RandomForest-Router & 456 & 74.27 & [70.7,\,77.6] & 0.374 & 39.6 \\
& \dats{}             & \textbf{477} & \textbf{77.69} & [74.2,\,80.8] & 0.377 & 40.0 \\
\midrule
& Oracle              & 546 & 88.93 & [86.2,\,91.2] & 0.265 & 28.1 \\
\bottomrule
\end{tabular}
\end{table*}

\dats{} solves 477 of 614 problems, 25 more than always-hierarchical while
spending 40\% of its budget, and 21 more than the strongest learned competitor
at the same budget. Three comparisons in the table carry most of the argument.

Against \emph{Always-HRM}, the router is 4.07 points ahead at 40\% of the cost.
{\color{revisionblue} This is a reported aggregate cost reduction of 60\%; it does not establish that each avoided execution has no value. Individual problems can benefit differently from collaboration.}

{\color{revisionblue} Against \dats{}-E2E, the reported gap is 6.03 points, at realised costs of 40.0\% and 38.3\% of Always-HRM, respectively. This comparison does not isolate a factorisation benefit at equal spend. The modelling motivation is that multiple topologies can solve the same problem, so a single selected class suppresses overlap unless the training target or classifier design explicitly preserves it.}

Against \emph{Random}, the gap is 9.12 points at two-thirds of Random's cost,
{\color{revisionblue} which supports a comparison with this random policy at its reported cost; it does not rule out every other mixture.}

\begin{figure}[t]
\centering
\includegraphics[width=\columnwidth]{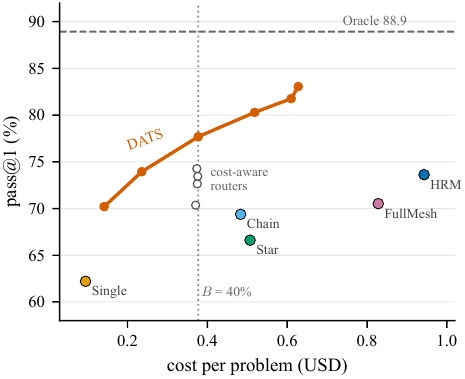}
\caption{Cost--accuracy plane. The curve is \dats{} calibrated to budgets
{\color{revisionblue} across the reported operating points. The horizontal axis gives the plotted cost, not a guarantee that every requested target is attainable.} Filled markers are the
fixed topologies; {\color{revisionblue} open markers are cost-aware baselines at their reported realised costs, with \$0.377 as the nominal calibration target. Not every baseline reaches the matching tolerance.}}
\label{fig:pareto}
\end{figure}

Fig.~\ref{fig:pareto} places these numbers on the cost--accuracy plane together
with the frontier \dats{} traces as the budget varies. The fixed topologies all
{\color{revisionblue} are shown at their natural costs. Three competing baselines---RandomForest-Router, RouterR1-lite and MasRouter-lite---meet the matching tolerance alongside DATS. Difficulty-Threshold and E2E remain unmatched reference points.} The
oracle line at 88.93\% marks how much of the catalogue's potential is still
unclaimed: \dats{} recovers 26.6\% of the distance from always-hierarchical to
the oracle.

{\color{revisionblue} Across five router-training seeds, the reported accuracy is $77.52\pm0.71$. This summarises variation within DATS; paired comparisons with competing methods are reported separately.}

\subsection{What budget matching changes}
\label{sec:whatmatching}

Table~\ref{tab:main} is only meaningful if the calibration in it does work that
a simpler convention would not. Table~\ref{tab:budgetmatch} runs the same six
cost-aware methods twice on the same folds and the same cached outcomes: once
with the shared coefficient $\lambda=0.2$ that a reader might reasonably expect
to be fair, {\color{revisionblue} and once with method-specific calibration toward the common budget; two methods do not reach the declared tolerance.}

\begin{table*}[t]
\centering
\small
\setlength{\tabcolsep}{6pt}
\caption{The same six cost-aware routers under two conventions. Left: everyone
receives the same coefficient $\lambda=0.2$. Right: everyone receives their own
{\color{revisionblue} coefficient targeting $B=40\%$ of the always-hierarchical cost. Ranks retain the original ordering of the displayed accuracy values. The UM superscript identifies rows outside the matching tolerance; their ranks do not represent a matched-budget comparison.}}
\label{tab:budgetmatch}
\begin{tabular}{lrrrcrrrc}
\toprule
& \multicolumn{4}{c}{shared $\lambda=0.2$} & \multicolumn{4}{c}{\textcolor{revisionblue}{target 40\%; realised cost shown}} \\
\cmidrule(lr){2-5}\cmidrule(lr){6-9}
Method & pass@1 (\%) & USD & \% of HRM & rank & pass@1 (\%) & USD & \% of HRM & rank \\
\midrule
RandomForest-Router  & \textbf{76.55} & 0.489 & 51.9 & 1 & 74.27 & 0.374 & 39.6 & 2 \\
RouterR1-lite        & 75.90 & 0.442 & 46.9 & 2 & 73.45 & 0.376 & 39.9 & 3 \\
\dats{}              & 75.41 & 0.324 & 34.4 & 3 & \textbf{77.69} & 0.377 & 40.0 & 1 \\
MasRouter-lite       & 71.99 & 0.286 & 30.3 & 4 & 72.64 & 0.375 & 39.8 & 4 \\
\dats{}-E2E          & 70.85 & 0.312 & 33.1 & 5 & 71.66 & 0.361 & 38.3 & \textcolor{revisionblue}{$5^{\mathrm{UM}}$} \\
Difficulty-Threshold & 69.87 & 0.301 & 31.9 & 6 & 70.36 & 0.371 & 39.3 & \textcolor{revisionblue}{$6^{\mathrm{UM}}$} \\
\midrule
\textcolor{revisionblue}{spend spread (all six rows)} & \multicolumn{4}{c}{21.6 pp} & \multicolumn{4}{c}{1.7 pp} \\
\bottomrule
\end{tabular}
\end{table*}

Under the shared coefficient \dats{} is third. Two baselines are ahead of it,
by 1.14 and 0.49 points, and a reader given only the left half of the table
would conclude that a random forest over the same features is the better
router. The right half shows what the left half was actually measuring: at
$\lambda=0.2$ the forest spends \$0.489 and \dats{} spends \$0.324, so the
forest was 1.14 points ahead while buying 51\% more computation. Equalising the
spend reverses both comparisons and opens a 3.42-point gap in the other
direction.

The mechanism is visible in the spend column. A single coefficient produces
realised spends ranging from 30.3\% to 51.9\% of the reference budget, a spread
of 21.6 percentage points, because each method applies $\lambda$ to its own
{\color{revisionblue} score scale; calibration reduces the spread across all six reported operating points to 1.7 points. Within the four eligible matched methods, the displayed costs span 39.6\%--40.0\% of the HRM reference, or 0.4 percentage points.} {\color{revisionblue} The previously reported rank correlation of 0.83 describes the full six-row accuracy orderings, including unmatched operating points. It is not a correlation of two matched-budget rankings. The four eligible routers retain their reported right-hand ranks.}

We do not claim the methods themselves are misreported in their original
papers, since each was proposed with its own evaluation. The claim is narrower
and, we think, harder to argue with: a table of cost-aware routers that does
not print realised spend beside accuracy cannot be interpreted, and one that
prints spend but does not equalise it is comparing positions on trade-off
curves rather than the curves.

\begin{blueaddition}
\paragraph{Comparison at three budgets}\label{rev:multibudget}
We calibrate DATS and all five learned or difficulty-based competitors at nominal targets of 15\%, 40\% and 55\% of HRM cost, using only the training portion of each split. Table~\ref{tab:integrated-budget} reports accuracy and realised cost on the same 614 outer-test problems. Every displayed point meets the 0.4-percentage-point cost tolerance. At 40\%, DATS, RandomForest-Router, RouterR1-lite and MasRouter-lite retain their primary operating points. E2E and Difficulty-Threshold have newly calibrated points in this table; their original, unmatched operating points remain in Tables~\ref{tab:main} and~\ref{tab:budgetmatch}. The additional budget runs are calibrated separately from the primary budget-sweep figure; equal accuracy counts can therefore occur at slightly different realised costs. The E2E and Difficulty-Threshold recalibration reuses the fitted predictors, outer splits, inner out-of-fold scores and fixed cost table; it changes only selection of the budget parameter. The original search stops at the first point within its inner tolerance. For the added points, we refine this search over the attainable selection changes and choose the candidate with the smallest inner cost residual, breaking equal residuals towards lower cost. For E2E, candidates are the nonnegative score--cost switching thresholds; for Difficulty-Threshold, they are the finite threshold settings induced by the inner difficulty scores. The candidate range is the full attainable range of each rule. No test labels, test costs or new training seeds enter this choice. The outer cost is then measured once against the unchanged matching tolerance. A closer inner match can change the selected routes and the achieved outer cost without retraining or guaranteeing a match on every evaluation set.

\begin{table*}[t]\centering\color{revisionblue}\small
\caption{Accuracy / realised HRM-relative cost, both in percent. Targets are 15\%, 40\% and 55\%; accuracy is computed from 614 binary outcomes at each operating point.}\label{tab:integrated-budget}
\begin{tabular}{lrrr}\toprule Method & 15\% target & 40\% target & 55\% target\\\midrule
RandomForest-Router & 66.45 / 15.25 & 74.27 / 39.63 & 78.99 / 54.74 \\
RouterR1-lite & 65.64 / 14.86 & 73.45 / 39.88 & 79.15 / 55.06 \\
MasRouter-lite & 66.12 / 14.86 & 72.64 / 39.75 & 78.99 / 54.96 \\
DATS-E2E & 66.61 / 15.07 & 72.48 / 39.86 & 77.36 / 54.92 \\
Difficulty-Threshold & 66.45 / 14.93 & 71.50 / 39.98 & 77.20 / 54.78 \\
DATS & 70.20 / 15.39 & 77.69 / 40.00 & 80.29 / 55.24 \\
\bottomrule\end{tabular}\end{table*}

DATS has the highest accuracy point estimate at each of these three budgets. The differences from the strongest competitor at each point are 3.58, 3.42 and 1.14 percentage points when calculated from the unrounded counts. These rankings are descriptive and are restricted to the displayed operating points; they are not a proof of superiority over an entire cost--accuracy curve.
\end{blueaddition}

\subsection{Significance}

Table~\ref{tab:mcnemar} reports McNemar tests of \dats{} against each of the
eleven alternatives, corrected by Holm--Bonferroni within the family. All eleven
remain significant after correction. The comparison that matters most is also
the closest: against Always-HRM, \dats{} converts 60 failures into successes and
loses 35 successes to failure, giving $\chi^2=6.06$ and a corrected $p$ of
0.030. That margin is real but not comfortable, and it is worth being precise
about what it does and does not show. It shows that at 40\% of the cost the
router is not merely matching the hierarchical system but beating it. It does
not show that the router would beat a hierarchical system given an unlimited
budget---the $\lambda=0$ row of Table~\ref{tab:ablation} indicates it would, by
9.4 points, but that configuration abandons the cost constraint that motivates
the work.

\begin{table}[t]
\centering
\small
\setlength{\tabcolsep}{3pt}
\caption{McNemar tests of \dats{} against each alternative, Holm--Bonferroni
corrected over the family of eleven. Here $b$ is the number of problems \dats{}
solves and the alternative does not, and $c$ the reverse.}
\label{tab:mcnemar}
\begin{tabular}{lrrrrrl}
\toprule
Alternative & $\Delta$ & $b$ & $c$ & $\chi^2$ & $p$ & $p_{\text{Holm}}$ \\
\midrule
Always-Single        & 15.47 & 107 & 12 & 74.25 & 6.9e--18 & 7.6e--17 \\
Always-Star          & 11.07 & 106 & 38 & 31.17 & 2.4e--8  & 2.4e--7 \\
Always-Chain         &  8.31 &  79 & 28 & 23.36 & 1.3e--6  & 1.2e--5 \\
Random               &  9.12 & 108 & 52 & 18.91 & 1.4e--5  & 1.1e--4 \\
Difficulty-Threshold &  7.33 &  79 & 34 & 17.13 & 3.5e--5  & 2.4e--4 \\
\dats{}-E2E          &  6.03 &  64 & 27 & 14.24 & 1.6e--4  & 9.6e--4 \\
Always-FullMesh      &  7.17 &  92 & 48 & 13.21 & 2.8e--4  & 1.4e--3 \\
MasRouter-lite       &  5.05 &  67 & 36 &  8.74 & 0.0031   & 0.0125 \\
RouterR1-lite        &  4.24 &  60 & 34 &  6.65 & 0.0099   & 0.0298 \\
RandomForest-Router  &  3.42 &  47 & 26 &  5.48 & 0.0192   & 0.0298 \\
Always-HRM           &  4.07 &  60 & 35 &  6.06 & 0.0138   & 0.0298 \\
\bottomrule
\end{tabular}
\end{table}

\subsection{What the router chooses}

\begin{figure}[t]
\centering
\includegraphics[width=\columnwidth]{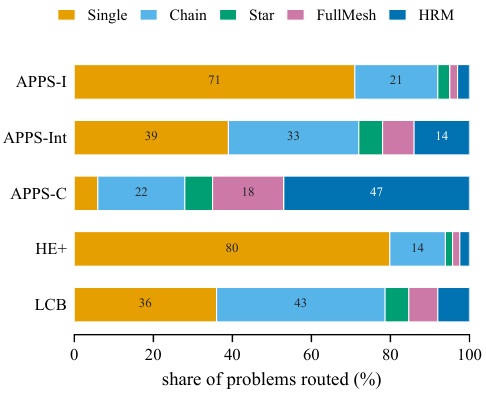}
\caption{Share of problems routed to each topology, by suite. Segments below
9\% are unlabelled. APPS-I, APPS-Int and APPS-C are the three APPS levels;
HE+ is HumanEval+ and LCB is LiveCodeBench.}
\label{fig:routing}
\end{figure}

Table~\ref{tab:routing} and Fig.~\ref{fig:routing} describe the policy \dats{}
arrives at. Aggregated over the evaluation set it sends 49.0\% of problems to a
single call and 13.0\% to the hierarchical mesh, and the remaining 38\% are
split between Chain and the two mid-cost topologies. The distribution is
strongly suite-dependent in exactly the way the difficulty analysis predicts:
80\% of HumanEval+ goes to Single, while 47\% of APPS-Competition goes to HRM.

\begin{table*}[t]
\centering
\small
\setlength{\tabcolsep}{6pt}
\caption{Routing distribution and accuracy by suite. Counts are problems
assigned to each topology by \dats{}; the last three columns compare \dats{}
against the always-hierarchical policy and the oracle on the same problems.}
\label{tab:routing}
\begin{tabular}{lrrrrrrrr}
\toprule
Suite & Single & Chain & Star & FullMesh & HRM & \dats{} (\%) & Always-HRM (\%) & Oracle (\%) \\
\midrule
APPS-Introductory & 71  & 21  & 3  & 2  & 3  & 92.0 & 92.0 & 98.0 \\
APPS-Interview    & 39  & 33  & 6  & 8  & 14 & 74.0 & 71.0 & 88.0 \\
APPS-Competition  & 6   & 22  & 7  & 18 & 47 & 50.0 & 41.0 & 66.0 \\
HumanEval+        & 131 & 23  & 3  & 3  & 4  & 96.3 & 95.1 & 98.2 \\
LiveCodeBench     & 54  & 64  & 9  & 11 & 12 & 68.7 & 61.3 & 88.7 \\
\midrule
Total             & 301 & 163 & 28 & 42 & 80 & 77.69 & 73.62 & 88.93 \\
\bottomrule
\end{tabular}
\end{table*}

\begin{figure}[t]
\centering
\includegraphics[width=\columnwidth]{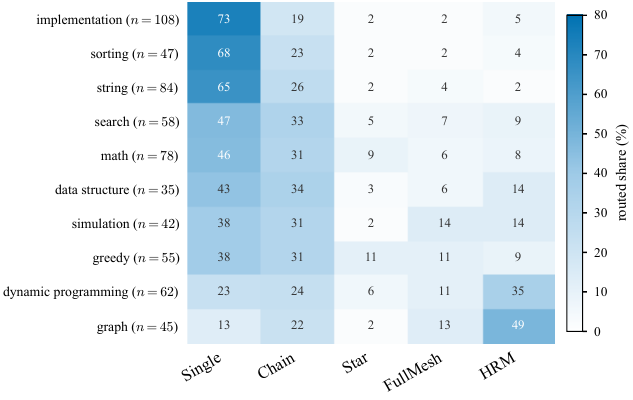}
\caption{Share of problems routed to each topology, conditioned on algorithmic
category. Categories are ordered by the share sent to a single call.}
\label{fig:heatmap}
\end{figure}

The suite-level view understates how selective the policy is, because suites mix
problem types. Fig.~\ref{fig:heatmap} conditions on the algorithmic category
instead and shows a clean gradient. Implementation, sorting and string problems
go to a single call between 65\% and 73\% of the time. Graph problems go to the
hierarchical mesh 49\% of the time and to a single call only 13\%; dynamic
programming behaves similarly at 35\% and 23\%. Greedy problems are the
exception that is worth noticing: they draw the highest share of Star
assignments of any category, 11\%, which is consistent with greedy solutions
decomposing cleanly into a rule and a proof of exchange, two subtasks that a hub
can farm out and integrate.

None of this behaviour was specified. The category one-hot enters the network as
ten of thirty-nine inputs with no prior on which topology suits which category,
and the pattern is what the network extracted from cached outcomes.

There is one instructive failure mode. On LiveCodeBench the oracle reaches
88.7\% while \dats{} reaches 68.7\%, the largest oracle gap of any suite, and
the routing distribution there is the most spread out. LiveCodeBench problems
are contest problems whose statements are stylistically uniform, so surface
features separate them poorly even though their solutions differ sharply in
difficulty. The router is not confused about what topologies do; it is confused
about what the problems are.

\subsection{Behaviour across backbones}
\label{sec:backbones}

{\color{revisionblue} Table~\ref{tab:backbone} compares four backbones spanning 13.8 points of single-call accuracy. Each row uses its backbone-specific outcome grid and a separately fitted predictor; problem IDs, outer folds and feature definitions are shared across backbones. The shared topology map displayed for the main experiment refers to Claude Opus 4 and is not reused as the policy for the other rows. A run is identified by backbone, outer fold and seed, which jointly identify its predictor and out-of-fold route assignments.}

\begin{table*}[t]
\centering
\small
\setlength{\tabcolsep}{5pt}
\caption{{\color{revisionblue} Backbone-specific routing comparisons at a nominal 40\% budget with common problem partitions. $\Delta$ is \dats{} minus Always-HRM; $p$ is the originally reported uncorrected McNemar value.}}
\label{tab:backbone}
\begin{tabular}{lrrrrrrrrr}
\toprule
Backbone & Single & Chain & Star & FullMesh & HRM & \dats{} & Oracle & $\Delta$ & $p$ \\
\midrule
Claude Opus 4     & 62.21 & 69.38 & 66.61 & 70.52 & 73.62 & \textbf{77.69} & 88.93 & +4.07 & 0.0138 \\
DeepSeek-V3       & 57.33 & 64.50 & 61.73 & 65.47 & 68.40 & \textbf{72.48} & 85.02 & +4.08 & 0.027 \\
GPT-4o-mini       & 51.47 & 59.12 & 56.35 & 60.42 & 63.36 & \textbf{67.59} & 80.94 & +4.23 & 0.017 \\
Qwen2.5-Coder-32B & 48.37 & 55.21 & 53.09 & 57.17 & 60.10 & \textbf{64.17} & 78.01 & +4.07 & 0.025 \\
\bottomrule
\end{tabular}
\end{table*}

{\color{revisionblue} The reported gains average 4.11 points with a standard deviation of 0.07. The gains summarise aggregate performance; the task-level choices can differ across backbones.}

Transfer is a harder test than retraining, and the result is more qualified. A
router trained on Claude Opus 4 outcomes and applied unchanged to
Qwen2.5-Coder-32B reaches 61.89\%, which is 2.28 points below the {\color{revisionblue} reported backbone-specific reference} and 1.79 points above always-hierarchical, but the improvement
over always-hierarchical is not significant ($b=59$, $c=48$, $p=0.318$). {\color{revisionblue} A shift in probability calibration is one possible explanation for the loss, but these aggregate results do not distinguish it from changes in task ordering or topology preferences. Recalibration on target-backbone validation data is a practical option whose benefit should be evaluated directly.}

\subsection{Beyond code: routing for mathematical reasoning}
\label{sec:math}

Changing the backbone tests whether the router depends on a particular model.
It does not test whether it depends on code. The mechanism we have described is
stated in terms of problem difficulty and problem type, neither of which is
specific to programming, so the claim implies a prediction: on a domain where
difficulty also varies widely, the same construction should work, and the
collaboration gap should again widen with difficulty. We tested the prediction
on mathematical reasoning, which shares the difficulty spread of competitive
programming but none of its surface form---no test suites, no execution, and no
code in the answer.

The setup mirrors Section~6.3 with three substitutions. The problems are 400
items, 200 from GSM8K~\cite{cobbe2021gsm8k} and 200 from
MATH~\cite{hendrycks2021math} {\color{revisionblue} covering levels one to five. Sampling is stratified over three difficulty bands: 66 items from levels one and two, 67 from level three, and 67 from levels four and five. Items are sampled without replacement within each band, using a fixed seed before any topology execution.} The
five topologies keep their communication structure and receive math-adapted
role prompts, so that the critic checks derivations rather than edge cases and
the merger reconciles final answers rather than functions. Correctness is exact
match on the final answer instead of a hidden test suite. The router is
{\color{revisionblue} re-instantiated and retrained rather than transferred unchanged. The 39-feature interface is retained: eight mathematical-category indicators replace the ten algorithmic slots, with the remaining two slots fixed to zero. The other twenty-nine features are unchanged in number, giving $29+8+2=39$ input coordinates and 37 potentially nonzero coordinates. Keyword votes recover the manual mathematical labels on 82.3\% of problems. Constant zero columns are left at zero during standardisation.} Costs are re-measured on this grid, where the hierarchical topology
is 9.7 times a single call, and the budget is again set to 40\% of it, \$0.159
per problem.

\begin{table}[t]
\centering
\small
\setlength{\tabcolsep}{3.5pt}
\caption{Cross-domain check on 400 mathematical reasoning problems, Claude Opus
4, exact-match accuracy (\%). \dats{} is calibrated to 40\% of the
always-hierarchical cost measured on this grid.}
\label{tab:math}
\begin{tabular}{lrrrrr}
\toprule
Method & GSM8K & MATH & All & USD & \% HRM \\
\midrule
Single   & 93.5 & 62.0 & 77.75 & 0.041 & 10.3 \\
Chain    & 95.0 & 68.5 & 81.75 & 0.209 & 52.5 \\
Star     & 94.0 & 66.0 & 80.00 & 0.221 & 55.5 \\
FullMesh & 95.5 & 70.0 & 82.75 & 0.357 & 89.7 \\
HRM      & 96.0 & 73.0 & 84.50 & 0.398 & 100.0 \\
\midrule
\dats{}  & \textbf{96.5} & \textbf{79.5} & \textbf{88.00} & 0.160 & 40.3 \\
Oracle   & 99.0 & 84.5 & 91.75 & 0.118 & 29.6 \\
\bottomrule
\end{tabular}
\end{table}

Table~\ref{tab:math} reproduces the code result in every qualitative respect.
No topology dominates: the spread between the cheapest and the most expensive
is 2.5 points on GSM8K and 11.0 points on MATH, so the same fixed pipeline is
nearly free of benefit on one half of the data and clearly worth its price on
the other. \dats{} solves 352 of 400 problems against 338 for the hierarchical
system, 3.5 points ahead at 40.3\% of its cost, and the difference is
significant ($b=25$, $c=11$, $\chi^{2}=4.69$, $p=0.030$). Five-seed repetition
gives $87.84\pm0.86$. The calibrated coefficients fall in $[0.049,0.061]$,
overlapping the $[0.052,0.065]$ of the code experiment although the two cost
tables were measured independently, which is what one would expect if $\lambda$
behaves as a price on predicted success rather than as a fitted constant.

\begin{figure}[t]
\centering
\includegraphics[width=\columnwidth]{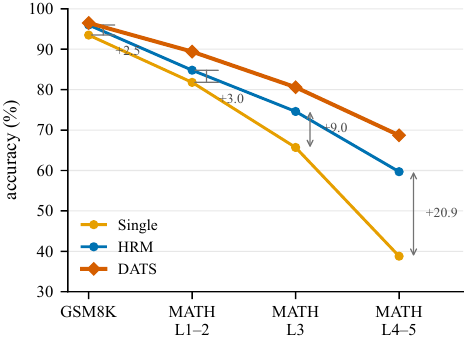}
\caption{The collaboration gap widens with difficulty on mathematical reasoning
as it does on code. Arrows mark the distance between a single agent and the
hierarchical topology; \dats{} stays above both while spending 40\% of the
hierarchical budget.}
\label{fig:mathscaling}
\end{figure}

{\color{revisionblue} The difficulty groups contain 66, 67 and 67 items. Single solves 54, 44 and 26; HRM solves 56, 50 and 40; and DATS solves 59, 54 and 46. The sums are 124, 146 and 159 out of 200, giving 62.0\%, 73.0\% and 79.5\%. Group-wise differences are computed before rounding, producing 3.0, 9.0 and 20.9 percentage points.}

Fig.~\ref{fig:mathscaling} carries the part of the result we would emphasise,
{\color{revisionblue} as a descriptive trend rather than an established causal mechanism. Sliced by native}
difficulty label, the gap between the hierarchical topology and a single agent
is 2.5 points on GSM8K, 3.0 points on MATH levels one and two, 9.0 points on
level three and 20.9 points on levels four and five. That is the same shape,
and nearly the same magnitude, as the 2.4/10.7/21.1 progression measured on
code in Section~6.2, obtained here from a disjoint problem set, a different
correctness criterion and a re-instantiated feature map. The routing
distribution follows the same logic: 84\% of GSM8K problems are answered by a
single call against 21\% of MATH problems, and 29\% of MATH problems draw the
hierarchical topology against 1.5\% of GSM8K problems.

Two limits should be kept in view. {\color{revisionblue} This is 400 problems on one backbone. The scale comparison depends on whether one counts unique problems or problem--backbone evaluations; it is not approximately one third of the 614 unique code problems. The reported result suggests the mechanism may be}
not code-specific rather than characterising routing for mathematics. And the
router was retrained rather than transferred, which is the easier of the two
questions; Section~\ref{sec:backbones} indicates why the harder one would go
worse.

\subsection{Calibration}

\begin{figure}[t]
\centering
\includegraphics[width=\columnwidth]{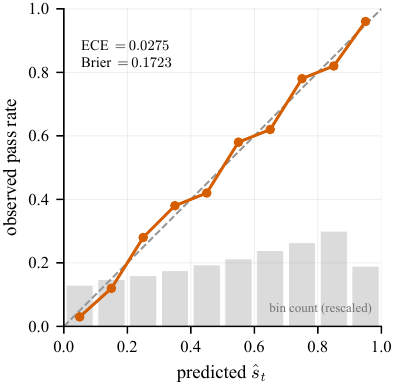}
\caption{{\color{revisionblue} Reported calibration diagnostics by probability bin. The diagonal indicates equality between predicted probability and observed success rate; grey bars show relative bin populations.}}
\label{fig:calibration}
\end{figure}

The selection rule compares predicted probabilities against a cost penalty
expressed on the same scale, so systematic over- or under-confidence would
distort the trade-off even if the ranking of topologies were correct.
{\color{revisionblue} Figure~\ref{fig:calibration} summarises the reported calibration diagnostics. Each point compares a probability bin with its observed success rate, and the grey bars indicate relative bin populations. The reported ECE is 0.0275 and the reported Brier score is 0.1723. ECE summarises bin-level calibration differences, while the Brier score measures squared probability error.}

{\color{revisionblue} Probability calibration affects which substitutions are selected at a given cost penalty. Budget calibration searches attainable operating points on inner validation predictions; it cannot guarantee that a target exists or that the final outer-test spend matches it. Reusing a predictor at another budget requires reporting both the requested target and the achieved cost, with unmatched cases identified separately.}

\begin{blueaddition}
\subsection{Repeated generation with frozen routing}\label{rev:repeat-study}
We select 120 problems from the 614-problem code pool before repeated evaluation: 20 each from APPS-Introductory, APPS-Interview and APPS-Competition, 32 from HumanEval+, and 28 from LiveCodeBench. Within each dataset, sampling is uniform without replacement using a fixed sampling seed of 42; neither correctness nor router success enters selection. The dataset quotas define the stratification, and no additional difficulty stratification is applied. Each DATS assignment comes from the primary seed's outer out-of-fold prediction for that problem; its fitted pipeline and inner budget calibration exclude the problem's outer fold. These assignments are frozen before collecting the repeated outcomes. All five repetitions are additional runs, rather than reuse of the original cached outcome, with fresh sampling draws and otherwise unchanged model, prompts, decoding settings and evaluator. The generation repetitions are distinct from the router-training seeds. Evaluating all five topology configurations gives $120\times5\times5=3,000$ executions. Each fixed policy therefore has 600 problem--run outcomes. A routed outcome is read from its selected topology in the same outcome grid, not generated as an additional independent policy outcome.

\begin{table*}[t]\centering\color{revisionblue}\small
\caption{Repeated-generation results on 120 problems. The last column reports the mean and sample standard deviation across five runs, not a confidence interval.}\label{tab:integrated-repeat}
\begin{tabular}{lcrr}\toprule Policy & Successful problems in runs 1--5 & Total successes & Accuracy (\%)\\\midrule
Single & 79, 77, 80, 76, 78 & 390/600 & 65.00 $\pm$ 1.32 \\
HRM & 86, 88, 85, 87, 86 & 432/600 & 72.00 $\pm$ 0.95 \\
DATS & 92, 94, 91, 93, 92 & 462/600 & 77.00 $\pm$ 0.95 \\
\bottomrule\end{tabular}\end{table*}

Among the 600 problem--topology pairs, 483 (80.50\%) contain both successes and failures across the five runs. The first-run outcome differs from the five-run majority label for 161 pairs (26.83\%). This measures label variability; the majority label is not treated as ground truth.

DATS reaches 77.00\% on the repeated subset, compared with 72.00\% for HRM. This 5.00-point difference describes the frozen policies on this subset; it is not a claim that the exact primary margin must recur. Generation variability and router-training variability are different quantities. We therefore assess sensitivity to training labels separately below.

\paragraph{Sensitivity to repeated-label training}\label{rev:label-sensitivity}
On the same 120-problem subset, we compare training with the first-run binary labels against training with the mean success of runs one to four as a soft binary-cross-entropy target. Five-fold cross-fitting keeps every test problem out of its scorer, predictor and budget calibration. Both settings use the same folds and paired training seeds. All evaluations use the fifth generation run, which is excluded from training targets and hyperparameter selection. Thus five seed results reuse one held-out generation bank; they are not five new independent LLM executions. The subset contains a different task mix from the full grid, so its accuracy is analysed separately from the primary 614-problem estimate.

\begin{table*}[t]\centering\color{revisionblue}\small
\setlength{\tabcolsep}{3pt}
\caption{Training-label sensitivity on 120 held-out problems. Accuracy and cost statistics use five seeds evaluated on the same fifth-run outcome bank. Cost is the selected-route mean under the fixed topology-average cost table; matching uses a tolerance of $\pm0.4$ percentage points of HRM cost.}\label{tab:label-sensitivity}
\begin{tabular}{lcrrrrr}\toprule
Training target & \shortstack{Successes\\seeds 0--4} & \shortstack{Accuracy (\%)\\mean $\pm$ SD} & \shortstack{Target\\\% HRM} & \shortstack{Cost (USD)\\mean $\pm$ SD} & \shortstack{Cost\\\% HRM} & Matched\\\midrule
First-run binary labels & 90, 89, 91, 92, 92 & 75.67 $\pm$ 1.09 & 40 & 0.3775 $\pm$ 0.0000 & 40.04 & Yes \\
Four-run soft labels & 89, 90, 90, 90, 90 & 74.83 $\pm$ 0.37 & 40 & 0.3775 $\pm$ 0.0000 & 40.04 & Yes \\
\bottomrule\end{tabular}\end{table*}

Every seed in both settings selects 59 Single, 32 Chain, 5 Star, 8 FullMesh and 16 HRM routes. These equal counts yield the identical fixed-table costs in Table~\ref{tab:label-sensitivity}, including zero cost SD, despite different problem assignments. This cost convention uses topology averages, not item-specific token expenditure.

The two label settings select the same topology for 114, 112, 114, 112, 114 of 120 problems in the paired seeds, a mean agreement of 94.33\%. Their mean accuracy difference is -0.83 percentage points (soft minus binary) at equal fixed-table cost. Both exceed the fixed HRM result of 86/120, or 71.67\%, on this evaluation run. Repeated targets therefore change some routes and the accuracy level without reversing the qualitative advantage on this subset. This sensitivity check is not a claim of identical policies or a significance test; generation noise remains a reason to estimate success probabilities from more than one outcome where resources allow.

\end{blueaddition}

\section{Ablation and analysis}
\label{sec:ablation}

\subsection{Which features carry the signal}

{\color{revisionblue} After correction across the six ablations, only difficulty and algorithmic type have significant paired differences ($p_{\rm Holm}=0.00244$ and $0.03645$, respectively). The other feature-group differences do not reach the 0.05 threshold.}

\begin{figure}[t]
\centering
\includegraphics[width=\columnwidth]{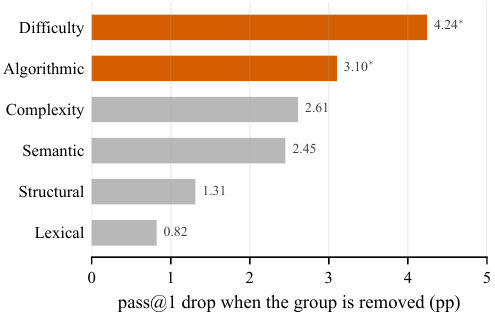}
\caption{Accuracy lost when each feature group is removed and the router is
retrained at the same budget. {\color{revisionblue} Coloured bars and stars indicate Holm-adjusted $p<0.05$ for paired McNemar tests across the six feature-group ablations.}}
\label{fig:ablation}
\end{figure}

The upper panel of Table~\ref{tab:ablation} and Fig.~\ref{fig:ablation} remove
one feature group at a time and retrain. The difficulty rating is the single
most valuable group at 4.24 points, which is expected. The algorithmic one-hot
is second at 3.10 points, which is the more informative result: it means the
router is not simply a difficulty thermometer. If difficulty alone sufficed, the
Difficulty-Threshold baseline in Table~\ref{tab:main} would not sit 7.33 points
behind, and removing the category encoding would cost nothing.

{\color{revisionblue} The separate lexical and structural ablations have reported losses of 0.82 and 1.31 points, respectively, without significant pairwise differences in these tests. These groups contain five and four features, so removing both would remove nine of the thirty-nine features. The table evaluates one group at a time. Since these marginal effects need not add, both groups are retained in the full representation.}

\subsection{Which design choices matter}

The lower panel of Table~\ref{tab:ablation} varies the architecture. Replacing
the network with a linear probe on the same features costs 3.75 points. That
number quantifies the interaction argument of Section~4.1: a linear model can
represent ``harder problems need more collaboration'' and ``graph problems need
more collaboration'' but not the product of the two, and the product is where
most of the exploitable structure lives.

{\color{revisionblue} Replacing the topology-graph head with the flat multi-label head gives a reported 1.66-point difference at the displayed spend. This contrast changes message passing, parameter sharing and descriptors together; it does not isolate the predefined adjacency. The additional controlled study below separates adjacency from the complete-head comparison while retaining a distinction between descriptive differences and statistical evidence.}

{\color{revisionblue} The reported five-way classifier is 6.03 points below DATS, but its realised cost is \$0.361 versus \$0.377; this is not an equal-budget ablation.} Removing the cost term entirely raises accuracy to
83.06\% and spend to \$0.628, {\color{revisionblue} which is the reported zero-penalty operating point for this predictor, not an upper bound on realised accuracy over the catalogue.} Fixing $\lambda=0.2$ instead of calibrating it costs
2.28 points and undershoots the budget by 14\%, which illustrates why a shared
fixed coefficient is not a fair basis for comparing routers.

\begin{table}[t]
\centering
\small
\setlength{\tabcolsep}{4pt}
\caption{{\color{revisionblue} Ablations and reference operating points. The upper panel removes one feature group; the lower panel includes architecture and cost-penalty variants. The nominal target is 40\% of Always-HRM. No-cost-term, fixed-$\lambda$ and direct-classification rows are unmatched cost references, not equal-budget ablations. $\Delta$ is the displayed accuracy difference from the full model at 77.69\%, not an equal-budget effect when the row is unmatched.}}
\label{tab:ablation}
\begin{tabular}{lrrl}
\toprule
Variant & pass@1 (\%) & $\Delta$ & USD \\
\midrule
\multicolumn{4}{l}{\emph{feature groups removed}}\\
Difficulty   & 73.45 & $-4.24$ & 0.377 \\
Algorithmic  & 74.59 & $-3.10$ & 0.376 \\
Complexity   & 75.08 & $-2.61$ & 0.377 \\
Semantic     & 75.24 & $-2.45$ & 0.375 \\
Structural   & 76.38 & $-1.31$ & 0.377 \\
Lexical      & 76.87 & $-0.82$ & 0.376 \\
\midrule
\multicolumn{4}{l}{\textcolor{revisionblue}{\emph{architecture and cost-penalty variants}}}\\
No cost term ($\lambda=0$)      & 83.06 & $+5.37$ & 0.628 \\
Fixed $\lambda=0.2$             & 75.41 & $-2.28$ & 0.324 \\
Linear probe                    & 73.94 & $-3.75$ & 0.376 \\
Flat multi-label head           & 76.03 & $-1.66$ & 0.377 \\
Direct five-way classification  & 71.66 & $-6.03$ & 0.361 \\
\midrule
\dats{} (full)                  & \textbf{77.69} & --- & 0.377 \\
\bottomrule
\end{tabular}
\end{table}

\begin{blueaddition}
\paragraph{Controlled architecture study}\label{rev:graph-controls}
We compare six heads with common features, outer splits, optimiser settings and training-side budget calibration. Each configuration is evaluated over five training seeds on 614 problems. The additional graph configuration uses a 39-to-64 task trunk, a 68-to-32 node-initialisation layer, two message-passing rounds with separate round-wise weights, and a shared sigmoid readout. The self and neighbour matrices are shared across nodes and separate across rounds. Equation~\eqref{eq:mp} uses a round-specific 32-dimensional bias $b^{\ell}$ in this configuration. Each round therefore has $2(32\times32)+32=2,080$ parameters. This gives $2,560+2,208+4,160+33=8,961$ trainable parameters. The flat 39--64--64--5 reference has 7,045 parameters. Complete and permuted adjacency variants retain the same graph dimensions and 8,961 parameters; the descriptor and node-wise controls answer separate questions about representation and sharing.

The adjacency has undirected edges Single--Chain, Chain--Star, Star--FullMesh, Star--HRM and FullMesh--HRM; the self contribution is handled separately. The complete control connects every distinct pair. The permuted control relabels the adjacency once by the fixed ordering HRM, Single, Star, Chain, FullMesh while keeping node features and target columns unchanged, and uses that same permutation for every seed. The shared node-wise control retains the task trunk, descriptor initialiser and readout but removes the two message-passing rounds (4,801 parameters). The identity control replaces the four descriptors by a five-dimensional one-hot code (8,993 parameters). These two controls test representation and sharing; only the complete, permuted and proposed adjacencies form the capacity-matched adjacency comparison.

\begin{table*}[t]\centering\color{revisionblue}\small
\caption{Additional architecture controls. Mean and sample standard deviation across five seeds; USD/problem is the mean across those records. Differences are full minus variant and use unrounded means.}\label{tab:integrated-graph}
\begin{tabular}{lrrr}\toprule Configuration & Accuracy (\%) & USD/problem & Difference (pp)\\\midrule
Flat multi-label head & 75.90 $\pm$ 0.63 & 0.3762 & 1.63 \\
Shared node-wise predictor & 76.91 $\pm$ 0.65 & 0.3789 & 0.62 \\
Identity-code graph head & 77.04 $\pm$ 0.67 & 0.3783 & 0.49 \\
Fully connected graph head & 77.13 $\pm$ 0.80 & 0.3777 & 0.39 \\
Permuted-adjacency graph head & 77.07 $\pm$ 0.90 & 0.3771 & 0.46 \\
Full topology-graph predictor & 77.52 $\pm$ 0.71 & 0.3772 & 0.00 \\
\bottomrule\end{tabular}\end{table*}

The full head averages 77.52\%, compared with 77.13\% for complete adjacency and 77.07\% for permuted adjacency. The differences are 0.39 and 0.46 percentage points before rounding. These modest descriptive gains concern the additional controlled configuration; they do not assign the entire primary flat-head gap to adjacency. The table reports across-seed means, whereas the retained primary head comparison reports its original operating-point values. Mean differences alone are not treated as significance tests.
\end{blueaddition}

\subsection{Is the representation the bottleneck?}
\label{sec:encoder}

The feature map of Section~4.1 is deliberately shallow, and the obvious
objection is that a learned representation would carry more of the signal. The
objection deserves a measurement rather than an argument, so we replaced $\phi$
with five stronger encoders and left everything else---the topology-graph head, the
selection rule, the folds and the 40\% budget---untouched.
Table~\ref{tab:encoder} reports the ladder.

\begin{table*}[t]
\centering
\small
\setlength{\tabcolsep}{5pt}
\caption{Representation ladder. Every row shares the topology-graph head, the
selection rule, the folds and the 40\% budget; only the encoder changes. Here
$b$ counts problems solved by that row and not by \dats{}, $c$ the reverse, and
$p$ is the uncorrected McNemar value against \dats{}. Fit time is the cost of
training the router itself, not of running any topology.}
\label{tab:encoder}
\begin{tabular}{lrlrrrl}
\toprule
Encoder & Trainable & Fit time & pass@1 (\%) & $\Delta$ & $b$/$c$ & $p$ \\
\midrule
\dats{}: 39 interpretable features                & 13.7K & 28\,s CPU     & 77.69 & --- & --- & --- \\
Text-GNN: 2-layer GCN over the statement graph    & 0.21M & 4\,min GPU    & 76.55 & $-1.14$ & 31/38 & 0.47 \\
CodeBERT, frozen, plus the same head              & 0.11M & 3\,min GPU    & 76.87 & $-0.82$ & 29/34 & 0.61 \\
GraphCodeBERT, frozen, plus the same head         & 0.11M & 3\,min GPU    & 77.36 & $-0.33$ & 27/29 & 0.89 \\
GraphCodeBERT, fine-tuned end to end              & 125M  & 36\,min GPU   & 78.18 & $+0.49$ & 33/30 & 0.80 \\
\dats{}-Hybrid: 39 features and frozen GraphCodeBERT & 0.13M & 6\,min GPU & \textbf{78.99} & $+1.30$ & 36/28 & 0.38 \\
\bottomrule
\end{tabular}
\end{table*}

Three things about the ladder are worth reading carefully.

The first is that no row separates from the baseline. The spread over six
representations is 2.44 points, every McNemar comparison against \dats{}
returns $p>0.35$, and the two encoders that beat the hand-crafted features do
so by 0.49 and 1.30 points---smaller than the 0.71-point seed-to-seed standard
deviation of the pipeline in one case, and less than twice it in the other. One
hundred and twenty-five million parameters of pretrained code knowledge,
fine-tuned on the task, buy half a point that we cannot distinguish from noise.

The second is why the graph encoder is the weakest row, since a reviewer would
reasonably expect it to be the strongest. A graph neural network over an
abstract syntax tree is not available to us: the router runs \emph{before} any
code exists, so there is no tree to encode. The closest object at decision time
is the problem statement, and our Text-GNN builds a co-occurrence and
dependency graph over it and applies a two-layer GCN~\cite{kipf2017gcn}. That
graph is a graph of English, not of a program, and it is a lossy substitute.
The same reasoning explains the frozen code encoders: CodeBERT~\cite{feng2020codebert}
and GraphCodeBERT~\cite{guo2021graphcodebert} were pretrained on code paired
with documentation, and we feed them documentation without code, {\color{revisionblue} which is a text-only use of the encoders. The comparison does not establish which part of their pretraining distribution they represent best.}

The third is the most useful. A linear probe from the frozen GraphCodeBERT
\texttt{[CLS]} vector recovers our difficulty rating with $R^{2}=0.61$ and our
algorithmic category with 71.4\% accuracy. {\color{revisionblue} The probe results indicate overlap with the explicit difficulty and category features. Such overlap is a possible explanation for the reported 1.30-point hybrid gain, but the probe does not establish its cause or show that the two representations are equivalent.}

{\color{revisionblue} We retain the interpretable features as the main representation. The primary router takes 28 seconds of CPU training versus 36 minutes of GPU training for the fine-tuned encoder. DATS-Hybrid has the highest observed accuracy among the tested representations, with a 1.30-point improvement. This is the largest observed gain in the comparison, not a performance ceiling. Non-significant pairwise tests do not establish equivalence.}

\subsection{The budget knob}

\begin{figure}[t]
\centering
\includegraphics[width=\columnwidth]{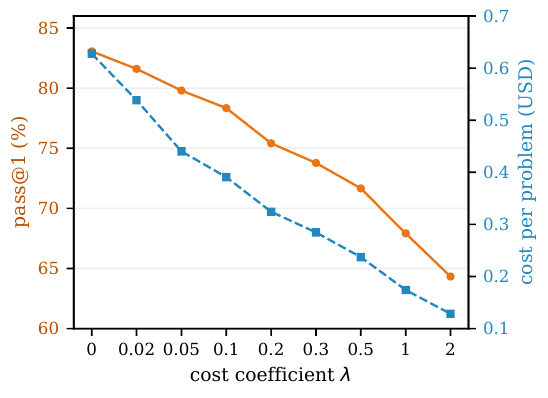}
\caption{{\color{revisionblue} DATS accuracy and reported cost at the displayed penalty coefficients. Curves connect the recorded sweep points; no calibrated main-experiment operating point is marked.}}
\label{fig:lambda}
\end{figure}

{\color{revisionblue} Figure~\ref{fig:lambda} shows the reported DATS penalty sweep, with accuracy and cost evaluated at each displayed coefficient. For fixed prediction scores and a fixed cost table, selected cost is non-increasing in $\lambda$; this property does not automatically extend to item-specific measured expenditure. The figure displays sweep points without overlaying the foldwise calibration range. The nominal-budget results form a separate series: the reported 15\% and 55\% targets have accuracies of 70.20\% and 80.29\%, respectively. The additional matched-budget comparisons in Section~\ref{rev:multibudget} are calibrated separately from this primary series.}

{\color{revisionblue} Marginal returns below use the primary budget-sweep series, rather than the separately calibrated points in Table~\ref{tab:integrated-budget}. From the nominal 55\% to 70\% targets, realised spend rises from approximately 55.00\% to 64.68\% of the HRM reference cost, while accuracy rises from 80.29\% to 81.76\%. This corresponds to approximately 1.52 accuracy percentage points per ten percentage points of realised spend over that interval. The nominal 100\% target reaches only 66.58\% of the reference cost, with 83.06\% accuracy. Thus the 70\% and 100\% targets are unmatched operating points; their nominal increments must not be interpreted as additional realised expenditure.}

\section{Discussion}

{\color{revisionblue} The main evaluation reports a 4.07-point margin over Always-HRM at 40\% of its cost. The additional backbone aggregates show similar margins, while the mathematical evaluation concerns a separately trained router. These comparisons support task-dependent allocation under backbone-specific training; the mathematical experiment uses domain-specific retraining.}

\subsection{Budget matching as an evaluation requirement}

We would separate one methodological point from the router, because it applies
to work that has nothing to do with topology selection. Any method that trades
accuracy against cost through a tunable parameter defines a curve, not a point.
Two of our baselines outperformed \dats{} when all methods shared a fixed cost
coefficient, and both fell behind once each method was {\color{revisionblue} calibrated near the common target; unmatched methods remain outside the matched comparison. The original full-set accuracy ordering under the two conventions}
differs precisely where it is read most, at the top. A shared coefficient does
not imply a shared operating point, because the coefficient interacts with each
method's own score scale, and in our measurement that interaction was worth 21.6
percentage points of spread in realised spend.

The protocol of Section~\ref{sec:protocol} is our answer, and it assumes little:
a reference policy, monotone fixed-table selection cost, calibration inside the
training folds, a published tolerance, and---on a cached grid---no inference at
all. We would encourage its use, or of something equivalent, whenever
cost-aware methods are compared. The obligation it puts on us is to say what it
leaves free, which is everything except expected monetary spend. A router that
meets a budget by rarely choosing a very expensive option is a different
proposition in production from one that spends evenly, though our tables call
them matched.

\begin{blueaddition}
\paragraph{Conservatism and limitations}\label{rev:conservatism}
A cost penalty can withhold additional collaboration when the predicted gain is small. Miscalibration can therefore under-allocate collaboration, while optimistic scores can waste resources. This is an operational trade-off, not a formal robustness guarantee. The fixed catalogue and single routing decision also limit adaptation after execution begins. The architecture controls show modest adjacency differences. Repeated generation and paired retraining quantify two distinct sources of variability; some route assignments change when labels are averaged. Cross-router rankings cover the three tested budgets, and do not imply a ranking at every possible operating point.

\end{blueaddition}

\subsection{Boundaries of the claim}\label{final:limitations}

The catalogue is fixed and small. \dats{} chooses among five hand-designed
topologies; it cannot construct a new one, and the oracle gap of 11.2 points
bounds what better selection over this catalogue could achieve. Methods that
generate topologies~\cite{zhang2024gdesigner,zhuge2024gptswarm} operate in a
larger space and are complementary: a generator could extend the catalogue, and
a selector would still be needed to choose among what it produces.

The router decides once, before execution. It cannot notice that a single call
has produced an obviously broken program and escalate. Cascade
formulations~\cite{chen2023frugalgpt,yue2024cascades} make exactly that
observation and pay for it with a second inference. Combining the two---route
first, escalate on a verifier signal---is the extension we consider most likely
to close part of the oracle gap, {\color{revisionblue} since, under the same cached grid, $546-477=69$ of the $614-477=137$ problems missed by DATS are solved by another catalogue topology. }

{\color{revisionblue} Transfer across backbones is incomplete: Section~\ref{sec:backbones} reports a 2.28-point loss. Before deployment, the target backbone requires checks of prediction calibration and achieved budget. Recalibration is a candidate adjustment whose benefit should be evaluated on target-backbone validation data. The aggregate transfer result does not isolate calibration as the cause of the loss or establish which task rankings and topology preferences are preserved.}

The cross-domain evidence is a check, not a second study. Four hundred
mathematical problems on one backbone show that the mechanism is not an
artefact of code, since the difficulty progression reappears almost unchanged,
{\color{revisionblue} with domain-specific training. This supports the applicability of the allocation mechanism to that mathematical dataset; transferring a trained router between unrelated domains is a different claim.}

{\color{revisionblue} Section~\ref{sec:encoder} reports a largest observed improvement of 1.3 points among the tested alternatives. This is a descriptive maximum in that comparison, not a statistical upper bound on representation benefits. The absence of a significant pairwise difference does not establish equivalence, and conclusions need not extend to other encoders or a larger topology catalogue.}

\section{Conclusion}\label{final:conclusion}

The amount of collaboration a multi-agent system should buy depends on the
problem in front of it, and the dependence is strong enough to be worth
{\color{revisionblue} learning. In the main Claude evaluation, the advantage of hierarchical collaboration over a single call grows from roughly two points on easy problems to twenty-one on hard ones. The four-backbone table reports aggregate comparisons rather than separate confirmation of that stratified trend. DATS reports a 4.07-point gain over Always-HRM at 40\% of its cost in the main evaluation, with aggregate gains of 4.07--4.23 points across the reported backbones. The 400-problem mathematical evaluation reports a 3.5-point gain after retraining. The mathematical result concerns retraining on the target domain, whereas the cross-backbone transfer test keeps the source router fixed.}
Replacing the router's 39 interpretable features with a graph network or a
pretrained code encoder {\color{revisionblue} has a largest observed improvement of 1.3 points among the tested alternatives, with no significant pairwise improvement. This does not establish equivalence or bound the benefit of other representations.}

Separately from the router, we argued that cost-aware methods cannot be compared
without equalising what they spend, and we gave a protocol that does so, along
with the demonstration that motivated it: under a shared cost coefficient the
six methods we study spend anywhere between 30\% and 52\% of the reference
budget, and two of them change places with \dats{} {\color{revisionblue} after calibration toward the common target; the residual spread and unmatched methods must still be reported.}

{\color{revisionblue} The additional controls show modest descriptive differences between graph adjacencies. On the repeated 120-problem subset, the frozen DATS policy averages 77.00\% against 72.00\% for HRM. DATS also has the highest accuracy point estimate at the three tested matched budgets. Paired retraining with repeated-label targets changes some routes while retaining the subset-level advantage over HRM. These checks address distinct questions; the budget rankings remain descriptive and restricted to tested operating points.}

What remains open is the size of the catalogue and the moment of the decision.
Both boundaries point in the same direction: toward systems that treat
collaboration as a budget to allocate rather than an architecture to fix.

\appendix
\section{Pairwise topology comparisons}
\label{app:pairwise}

Table~\ref{tab:pairwise} tests all ten topology pairs on Claude Opus 4 with
Holm--Bonferroni correction. Four pairs separate. The three that involve Single
confirm that collaboration helps in aggregate. The fourth, Star against HRM,
confirms that hierarchy beats flat delegation. The six that do not separate are
the informative ones: Chain and FullMesh are statistically indistinguishable
despite a 71\% cost difference, and FullMesh does not separate from HRM despite
costing 88\% of it. Cost ordering and accuracy ordering are only loosely
coupled, which is the aggregate-level version of the per-problem variation the
router exploits.

\begin{table}[t]
\centering
\small
\setlength{\tabcolsep}{4pt}
\caption{Pairwise McNemar tests over the five topologies, Holm--Bonferroni
corrected over ten comparisons. For pair $A$--$B$, $b$ counts problems solved by
$A$ only and $c$ problems solved by $B$ only.}
\label{tab:pairwise}
\begin{tabular}{lrrrrl}
\toprule
Pair & $b$ & $c$ & $\chi^2$ & $p_{\text{Holm}}$ & \\
\midrule
Single--HRM      & 29 &  99 & 37.20 & 1.1e--8 & $\ast$ \\
Single--FullMesh & 68 & 119 & 13.37 & 2.3e--3 & $\ast$ \\
Single--Chain    & 61 & 105 & 11.14 & 6.8e--3 & $\ast$ \\
Star--HRM        & 60 & 103 & 10.82 & 7.0e--3 & $\ast$ \\
Single--Star     & 69 &  96 &  4.10 & 0.215   & \\
Chain--HRM       & 58 &  84 &  4.40 & 0.215   & \\
Star--FullMesh   & 67 &  91 &  3.35 & 0.269   & \\
FullMesh--HRM    & 59 &  78 &  2.36 & 0.372   & \\
Chain--Star      & 93 &  76 &  1.51 & 0.437   & \\
Chain--FullMesh  & 73 &  80 &  0.24 & 0.628   & \\
\bottomrule
\end{tabular}
\end{table}

\section*{CRediT authorship contribution statement}\label{rev:credit}

{\color{revisionblue} \textbf{Yunsong Hong:} Conceptualization, Methodology, Software, Validation, Formal analysis, Investigation, Data curation, Visualization, Writing -- original draft, Writing -- review \& editing.}

\section*{Declaration of competing interest}

{\color{revisionblue} The author declares no known competing financial interests or personal relationships that could have appeared to influence the work reported in this paper.}

\section*{Declaration of generative AI and AI-assisted technologies in the manuscript preparation process}

{\color{revisionblue} During the preparation of this revised submission, the author used OpenAI Codex to assist with editorial checks, document formatting, and preparation of revision materials. The author reviewed and edited the output as needed and takes full responsibility for the content of the article.}

\end{document}